\documentclass[journal]{new-aiaa}
\usepackage[utf8]{inputenc}
\usepackage{textcomp}

\usepackage{graphicx}
\usepackage{amsmath}
\usepackage[version=4]{mhchem}
\usepackage{siunitx}
\usepackage{longtable,tabularx}
\usepackage{footnpag}			     

\usepackage{amssymb}
\usepackage{amsbsy}
\usepackage{float}
\usepackage{booktabs}
\usepackage{caption}
\usepackage{subcaption}
\usepackage[capitalise]{cleveref}
\usepackage{longtable}
\usepackage[
    starfontserif 
    ]{starfont}
\DeclareSymbolFont{starfontsym}{OT1}{sts}{m}{n}
\DeclareMathSymbol{\mathUranus}{\mathord}{starfontsym}{70}
\DeclareMathSymbol{\mathvarUranus}{\mathord}{starfontsym}{65}
\newcommand{\mbf}{\mathbf}
\newcommand{\xvec}{\mbf{x}}

\allowdisplaybreaks

\title{Improved Directional State Transition Tensors for Accurate Aerocapture Performance Analysis}

\author{Grace E. Calkins\footnote{Ph.D. Pre-Candidate, Department of Aerospace Engineering Sciences. grace.calkins@colorado.edu} and Jay W. McMahon,\footnote{Associate Professor, Department of Aerospace Engineering Sciences.}}
\affil{University of Colorado Boulder, Boulder, CO 80303}
\author{David C. Woffinden\footnote{Aerospace Engineer, GN\&C Autonomous Systems Branch}}
\affil{NASA Johnson Space Center, Houston, TX, 77508}

\begin{document}

\maketitle

\begin{abstract}
Aerocapture is particularly challenging for semi-analytical propagation because the dynamics are dominated by nonconservative forces whose magnitudes vary significantly throughout the trajectory. State transition tensors (STTs), higher-order Taylor series expansions of the solution flow, have been widely used as a computationally efficient semi-analytical propagation method for orbital scenarios, but have not previously been applied to aerocapture. However, computing higher-order STTs requires integrating exponentially many equations as the state dimension increases. Directional state transition tensors (DSTTs) mitigate this cost by projecting the state into a reduced-dimension basis. This work develops novel dynamics analysis techniques to identify effective bases for this reduction, including augmented higher-order Cauchy Green tensors tailored to quantities of interest such as apoapsis radius. Results show that DSTTs constructed along these bases significantly reduce computational cost while maintaining accuracy in predicted apoapsis radius and terminal energy. In particular, certain of these DSTTs outperform traditional DSTTs in nonlinear perturbation propagation for key state subsets and quantities of interest. These results establish STTs and DSTTs as practical tools for aerocapture performance analysis to enable robust guidance and navigation.
\end{abstract}

\section*{Nomenclature}

{\renewcommand\arraystretch{1.0}
\noindent
\begin{longtable*}{@{}l @{\quad=\quad} l@{}} 
	$\mathbf{A}$ & Dynamics partials \\ 
    $\boldsymbol{\mathcal{C}}^{[p]}$ & $p^{\text{th}}$-order Cauchy Green tensor \\
    $D$ & Drag acceleration [m/s$^2$] \\
    $\mathbf{D}_{[p]}$ & $p^{\text{th}}$-order coefficient of the quantity-of-interest Taylor series expansion \\
    $H$ & Exponential atmosphere scale height [m] \\
    $J_2$ & Second zonal harmonic \\
    $L$ & Lift acceleration [m/s$^2$] \\
    $\frac{L}{D}$ & Lift-to-drag ratio \\ 
    $\mathbf{Q}$ & QR decomposition matrix \\ 
    $\mathcal{Q}^{[p]}$ & $p^{\text{th}}$-order quantity-of-interest Cauchy Green tensor\\
    $\tilde{\mathcal{Q}}^{[p]}$ & $p^{\text{th}}$-order non-symmetric quantity-of-interest Cauchy Green tensor  \\
    $\mathbf{R}_{[p]}$ & $p^{\text{th}}$-order DSTT rotation matrix \\ 
    $\mathbf{R}_{\text{QR}}$ & QR decomposition matrix \\ 
    $R_{\mathvarUranus}$ & Radius of Uranus [m] \\
    $\mathbf{S}$ & Selection matrix \\
    $\mathcal{S}^{[p]}$ & $p^{\text{th}}$-order selective Cauchy Green tensor\\
    $\tilde{\mathcal{S}}^{[p]}$ & $p^{\text{th}}$-order non-symmetric selective Cauchy Green tensor  \\
    $V$ & Planet-relative velocity [m/s] \\
    $V_{\infty}$ & Hyperbolic excess velocity [m/s] \\
    $a$ & Semi-major axis [m] \\ 
    $a_{\text{ratio}}$ & Ratio of aerodynamic to gravitational accelerations \\ 
    $\mathbf{d}$ & Column of $\mathbf{Q}$ matrix \\ 
    $\mathbf{e}$ & Standard basis vector \\ 
    $f$ & Nonlinear dynamics function \\ 
    $g_\phi$ & Latitudinal component of gravitational acceleration [m/s$^2$] \\
    $g_r$ & Radial component of gravitational acceleration [m/s$^2$] \\
    $h$ & Altitude [m] \\ 
    $h_0$ & Exponential atmosphere reference height [m] \\
    $l$ & Number of sensitive directions used for DSTT construction \\ 
    $m_{\text{ref}}$ & Reference mass [kg] \\ 
    $n$ & State dimension \\ 
    $p$ & STT/DSTT order \\
    $p_d$ & Dynamic pressure [Pa] \\
    $q$ & Quantity of interest function \\
    $\mathbf{q}$ & Quantity of interest  \\
    $r$ & Radial distance from the center of the planet [m] \\
    $r_a$ & Apoapsis radius [m] \\
    $t$ & Time [s] \\
    $\mathbf{u}$ & Vector orthogonal to $\mathbf{R}_{[2]}$ \\
    $\xvec$ & Vehicle state vector \\
    $\bar{\xvec}$ & Reference state \\
    $\mathbf{y}$ & Reduced-dimension state vector for DSTTs \\
    $\boldsymbol{\Phi}$ & State transition matrix \\
    $\tilde{\Phi}^{[p]}$ & Non-symmetric Cauchy Green Tensor of order $p$  \\
    $\Omega$ & Planet's constant angular velocity [$^\circ$/s] \\ 
    $\beta$ & Vehicle ballistic coefficient [kg/m$^2$] \\ 
    $\gamma$ & Flight path angle \\
    $\varepsilon$ & Specific orbital energy [m$^2$/s$^2$] \\ 
    $\epsilon_{\mathcal{F}}$ & Frobenius norm error \\ 
    $\epsilon_{r_a}$ & Predicted apoapsis radius error [m] \\ 
    $\epsilon_{\varepsilon}$ & Predicted specific energy error [m$^2$/s$^2$] \\ 
    $\zeta$ & Natural log of density \\ 
    $\eta^{i,\ldots}$ & Partial of quantity of interest with respect to the state \\
    $\theta$ & Latitude [$^\circ$] \\
    $\lambda$ & Eigenvalue \\ 
    $\mu$ & Planet's gravitational parameter [m$^3$/s$^2$] \\
    $\boldsymbol{\xi}$ & Selected states \\ 
    $\rho$ & Atmospheric density [kg/m$^3$] \\
    $\rho_0$ & Exponential atmosphere reference density [kg/m$^3$] \\
    $\sigma$ & Bank angle [$^\circ$] \\
    $\phi$ & Longitude [$^\circ$] \\
    $\boldsymbol{\phi}$ & Solution flow \\
    $\phi^{i, \gamma_1 \ldots \gamma_{p-1}}$, $\phi^{[p]}$  & $p^{\text{th}}$-order state transition tensor \\
    $\psi$ & Heading angle \\
    $\psi_{[p]}^{i,\ldots}$ & $p^{\text{th}}$-order DSTT element \\
\end{longtable*}}
\setcounter{table}{0}

\section{Introduction}

\lettrine{A}{erocapture} is an alternative to propulsive orbit insertion in which the spacecraft flies through the atmosphere of a planet to capture into orbit, rather than performing a large propulsive maneuver~\cite{spilker_qualitative_2019,girija_quantitative_2022}. As evidenced by the Uranus Orbiter Probe mission, aerocapture is an enabling technology for outer planet missions~\cite{dutta_uranus_2024}. While aerocapture can reduce propellant mass and transit time for such missions, aerocapture trajectories are characterized by inherent nonlinearities and potentially large deviations from the desired state. These attributes prove challenging for efficient performance analysis and uncertainty propagation schemes as they necessitate robust nonlinear methods. 

This paper first seeks to determine whether state transition tensors (STTs), a computationally efficient semi-analytical uncertainty propagation method previously applied to orbital scenarios, provide similar benefits when applied to nonconservative aerocapture dynamics.  Building on this, we aim to identify whether the complexity of the STTs can be reduced through directional state transition tensors (DSTTs) for aerocapture. More specifically, the STTs are a higher-order Taylor series expansion of the trajectory dynamics, while DSTTs provide a reduced-dimension representation of the STTs along a dynamically-important direction~\cite{park_nonlinear_2007,park_nonlinear_2006,boone_directional_2023}. While STTs require tensor contraction for nonlinear perturbation and moment propagation, DSTTs simplify these processes to matrix-vector multiplication for a single latent dimension.

This makes DSTTs advantageous compared to other common uncertainty quantification methods. Although Monte Carlo simulation is a prevalent nonlinear uncertainty quantification method, its high computational cost makes it infeasible for onboard implementation. To incorporate uncertainty in onboard applications\textemdash such as stochastic guidance algorithms or nonlinear filtering\textemdash there is a need for more computationally efficient methods that can still capture large initial error distributions and the inherent nonlinearities of the problem. Previous studies have either sought to linearize the aerocapture problem to perform efficient uncertainty propagation or apply computationally efficient nonlinear propagation schemes. To use linear propagation schemes, the nonlinear aerocapture equations of motion can be linearized directly and uncertainty can be propagated using Linear Covariance Analysis (LCA)~\cite{ridderhof_linear_2022,joshi_end-to-end_2022}, or the dynamics can be transformed into a set of coordinates known as quasi-initial conditions (QICs) which behave more linearly~\cite{grace_quasi-initial_2022}. Nonlinear uncertainty propagation methods, such as Polynomial Chaos Expansion (PCE)~\cite{grace_two-stage_2022,albert_finite-dimensional_2021}, and the Unscented Transform (UT)~\cite{grace_application_2020}, have also been explored for aerocapture uncertainty propagation. 

However, these methods\textemdash both linear and nonlinear\textemdash have a few limitations. QICs fail when entry uncertainties are large or when complex atmospheric models are utilized. Because the covariance dynamics are linearized, LCA cannot adequately capture nonlinear effects of the aerocapture dynamics in covariance propagation. PCE suffers from the curse of dimensionality, and thus cannot be used when a greater number of states, such as atmospheric or navigation parameters, need to be dispersed. While the UT can estimate mean and covariance of a Gaussian distribution to higher-order, its computational cost increases with the state dimension, $n$. The UT requires $2n+1$ numerical integrations of the nonlinear dynamics model at each simulation timestep. Compared to these methods, STTs and DSTTs are beneficial not only because they enable efficient uncertainty propagation, but because they can be used to directly compute relevant quantities (such as the nonlinearity metrics for Gaussian mixture model construction and guidance presented in \cite{kulik_applications_2024,kulik_nonlinearity_2024}).

As mentioned above, DSTTs offer a more efficient representation of nonlinear dynamics than standard STTs as they require fewer terms. Because of this, fewer equations must be integrated to obtain DSTTs than STTs. However, the greatest challenge in using DSTTs is selecting the dynamically important direction along which to reduce the basis dimension. The original DSTT formulation used the dominant eigenvector of the second-order Cauchy-Green tensor (CGT), the direction of maximum stretching in the linear dynamics, to construct DSTTs for orbital scenarios~\cite{boone_directional_2023}. The CGT is computed as the state transition matrix transpose times itself, and its eigenpairs identify the principal directions of expansion for the dynamical flow~\cite{guzzetti_stationkeeping_2017,short_mode_2017,oguri_convex_2019}.  However, the authors~\cite{calkins_grace_e_efficient_2024,calkins2025dynamics} found in previous work that directionalizing the DSTTs along the second-order CGT dominant eigenvector\textemdash as had been done in the original DSTT work on orbital scenarios~\cite{boone_directional_2023,boodram_efficient_2022}\textemdash did not result in accurate propagation for aerocapture.

Other work has explored alternative DSTT directionalization schemes. Recently, Zhou et al. developed a method in which they map a time-varying second-order dominant CGT eigenpair without needing to integrate the STTs~\cite{zhou_time-varying_2024}. This eliminates the need to compute the second-order CGT stretching directions in advance of constructing the DSTTs. However, this method necessitates identifying the dominant eigenvalues a priori. If the relative eigenvalue magnitudes change throughout the trajectory (which can happen in nonconservative systems), and the proper eigenpairs are not selected, the DSTT could no longer be an accurate approximation of the original STT. Thus, it is prudent to use multiple latent dimensions with this method to ensure that the maximal eigenvalue is always incorporated into the DSTT construction. This means a key benefit of DSTTs, only requiring matrix-vector multiplication rather than tensor contraction when using a single latent dimension, is lost. In addition, the time-varying DSTT method uses the eigenpairs of the second-order CGT to construct DSTTs, meaning only stretching due to the linear dynamics is considered. While this method is beneficial for reducing the number of equations to compute DSTTs onboard, it does not consider higher-order dynamics when constructing DSTTs. In the present work, we construct DSTTs along directions associated with nonlinear deformation a priori, rather than computing a time-varying direction along-the-way. 

Thus, this work improves DSTT performance for nonlinear dynamics, such as aerocapture, by addressing key limitations in the current theory. In previous work, the authors~\cite{calkins_grace_e_efficient_2024,calkins2025dynamics} and Zhou et al.~\cite{zhou_efficient_2025} constructed DSTTs considering an important factor in the application rather than the direction of linear stretching. These studies constructed DSTTs using a heuristic direction scaled by dynamic pressure for aerocapture and directions informed by uncertainty for orbit determination. In the present work, we seek to find dynamically-sensitive directions from the STTs to construct accurate DSTTs by developing a general method to find a higher-order sensitive directions in the dynamics. The proposed methodology can be applied to any nonconservative, nonlinear system. 

Additionally, this paper demonstrates that previous conclusions about DSTT construction are not true for aerocapture. Prior work on orbital dynamics found that (1) the maximal second-order CGT eigenvector direction (the maximum linear stretching direction) is relatively constant throughout the trajectory~\cite{boone_directional_2023,boodram_efficient_2022} and (2) the higher-order stretching directions align with the linear directions~\cite{boone_directional_2023,boone_efficient_2023}. Regarding the first observation, we demonstrate that nonconservative aerodynamic forces result in time-varying stretching direction for aerocapture dynamics, rather than the approximately constant direction observed in orbital dynamics. Additionally, as the linear stretching direction does not align with the higher-order stretching direction, we construct improved DSTTs considering sensitive directions found using tensor eigenpairs of novel augmented higher-order Cauchy Green Tensors (HOCGTs). These tensor eigenpairs are the higher-order analog to the matrix eigenproblem and have been shown to identify directions of strong nonlinearity~\cite{jenson_semianalytical_2023}. The augmented HOCGTs identify directions of maximum stretching for a selected subset of the state or a quantity-of-interest that is an arbitrary function of the state. 

This paper is organized as follows. First, the aerocapture problem is presented in \cref{sec:aerocapture}. Then, STTs and DSTTs are introduced in \cref{sec:sttsdstts}. Third, dynamics analysis techniques using functions of the STTs are presented in \cref{sec:dynamics_analysis}, including decomposed CGTs showing the time-varying direction and augmented HOCGTs. In \cref{sec:results}, results are presented showing STT sufficiency for aerocapture perturbation propagation, DSTT approximation accuracy, and DSTT performance for propagating key quantities of interest. Finally, conclusions are given in \cref{sec:conclusions}.

\section{The Aerocapture Problem}
\label{sec:aerocapture}

The vehicle state is $\xvec = [r, \theta, \phi, V, \gamma, \psi, \zeta]^{\top}$, where $r$ is the radial distance from the center of the planet, $\theta$ and $\phi$ are the latitude and longitude, $V$ is the planet-relative velocity, $\gamma$ is the flight path angle, $\psi$ is the heading angle, and $\zeta=\ln(\rho)$ is the natural log of density. Although density is not commonly included as a state variable, it is appended to the state for this study to include atmosphere variation in state perturbations. The 3DOF equations of motion for the spacecraft inside the atmosphere of a rotating planet are~\cite{vihn_hypersonic_1980, lu_optimal_2015}:
\begin{align} \label{eqn:position}
	\dot{r}&=  V \sin \gamma, \\ 
	\dot{\theta}&=  \frac{V \cos \gamma \sin \psi}{r \cos \phi},\\
	\dot{\phi}&=  \frac{V \cos \gamma \cos \psi }{ r},\\
	\dot{V}&=  -D- g_r \sin \gamma -g_\phi \cos{\gamma}\cos{\phi}+\Omega^2 r \cos \phi (\sin \gamma \cos \phi - \cos \gamma \sin \phi \cos \psi),\\
	\dot{\gamma}&=\frac{1}{V}\left[L \cos \sigma+\left(V^2 / r-g_r\right) \cos \gamma+g_\phi \sin \gamma \cos \psi+2 \Omega V \cos \phi \sin \psi\right. \\
		&\left.+\Omega^2 r \cos \phi(\cos \gamma \cos \phi+\sin \gamma \cos \psi \sin \phi)\right] , \nonumber \\
	\dot{\psi}&=\frac{1}{V}\left[\frac{L \sin \sigma}{\cos \gamma}+\frac{V^2}{r} \cos \gamma \sin \psi \tan \phi+g_\phi \frac{\sin \psi}{\cos \gamma}-2 \Omega V(\tan \gamma \cos \psi \cos \phi-\sin \phi)\right. \\
		&\left.+\frac{\Omega^2 r}{\cos \gamma} \sin \psi \sin \phi \cos \phi\right], \qquad \text{and} \nonumber \\
 	\label{eqn:density}
	\dot{\zeta} &= -\frac{V \sin\gamma}{H},	
\end{align}
where $\sigma$ is the bank angle, $L$ and $D$ are the lift and drag accelerations, $\Omega$ is the planet's constant angular velocity, and $H$ is the exponential atmosphere scale height.\footnote{Although previous STT and DSTT studies employed Cartesian coordinates for orbital dynamics, Grace et al. demonstrated that spherical coordinates provide a more favorable representation for aerocapture uncertainty propagation than Cartesian coordinates~\cite{grace_quasi-initial_2022}.} These equations assume that the planet is spherical and has a constant angular velocity. 

 The equations to compute the latitudinal and longitudinal components of the gravity vector, $g_\theta$ and $g_\phi$, are:
\begin{align}
	g_r&=\frac{\mu}{r^2}\left[1+J_2\left(\frac{R_{\mathvarUranus}}{r}\right)^2\left(1.5-4.5 \sin ^2 \phi\right)\right] \\
	g_\phi &=\frac{\mu}{r^2}\left[J_2\left(\frac{R_{\mathvarUranus}}{r}\right)^2(3 \sin \phi \cos \phi)\right]
\end{align}
where $\mu$ is the planet's gravitational parameter, $R_{\mathvarUranus}$ is the radius of Uranus,  and $J_2$ is the second zonal harmonic. Lift and drag accelerations are computed as $L =  \frac{1}{2} \rho V^2 \frac{L}{D} \beta $ and  $D =  \frac{1}{2} \rho V^2 \beta$, where $\frac{L}{D}$ is the lift-to-drag ratio and $\beta = \frac{m}{C_D A_\text{ref}}$ is the ballistic coefficient. The vehicle mass is given by $m$, $C_D$ is the vehicle's drag coefficient, and $A_\text{ref}$ is the reference surface area of the vehicle. 

Given the sensitivity of the dynamics to atmospheric density and the scale differences between states, some considerations were taken to improve the numeric stability of the equations of motion. To address the significant variation in density along the trajectory (approximately five orders of magnitude), the natural logarithm of density is used as a state variable to reduce sensitivity in the partial derivatives used to compute the STTs. This equation of motion was derived assuming an exponential atmosphere model $\rho = \rho_0 \exp{\frac{h_0 -(r - R_{\mathvarUranus})}{H}}$, where the reference density $\rho_0=6.40\times 10^{-3}$ kg/m$^3$, the reference height $h_0=0$ km, and the scale height $H=54.72$ km. The exponential atmosphere constants were found by performing a fit to the mean of data from UranusGRAM with a density perturbation scale of two~\cite{justh_uranus_2021}. To further improve numeric stability, the nondimensionalization scheme developed by Lu was implemented \cite{lu_entry_2014}. The nondimensional ODEs are equivalent to the dimensional ODEs in Eqs.~(\ref{eqn:position}-\ref{eqn:density}), except that all dimensional variables are replaced with their nondimensional counterparts. A natural log of density normalization factor $\zeta_\text{ref}=20$ was also introduced, and the natural log of density is normalized as $\zeta^* = \zeta /\zeta_\text{ref}$.
 
The initial state for the results presented in this paper is given in \cref{tab:init_state}, where the initial radial position is given as an altitude such that $r_0 = h_0 + R_{\mathvarUranus}$. These values are chosen to be similar to the proposed Flagship-class Uranus Orbiter Probe (UOP) mission, which targets a highly elliptical orbit with an apoapsis altitude of 550,000 km and periapsis altitude of 4,000 km \cite{deshmukh_performance_2024}. All trajectories are simulated for 780 seconds to encompass the atmospheric flight duration.  Two entry velocities, corresponding to 13.6 km/s and 17.5 km/s $V_{\infty}$, are considered~\cite{deshmukh_performance_2024,restrepo_mission_2024}. The entry flight path angle for each scenario was selected as the center of its entry flight path angle corridor, and the constant bank angle was chosen to achieve the target apoapsis radius. 

\begin{table}[h!]
\centering
\caption{Initial State. Initial velocity and flight path angle are in the inertial frame.}
\label{tab:init_state}
\begin{tabular}{lcccccccccc}
    \hline
     Scenario & 
\begin{tabular}{@{}c@{}}$h$ \\ $[$km$]$\end{tabular} & 
\begin{tabular}{@{}c@{}}$\theta$ \\ $[^\circ]$\end{tabular} & 
\begin{tabular}{@{}c@{}}$\phi$ \\ $[^\circ]$\end{tabular} & 
\begin{tabular}{@{}c@{}}$V$ \\ $[$km/s$]$\end{tabular} & 
\begin{tabular}{@{}c@{}}$\gamma$ \\ $[^\circ]$\end{tabular} & 
\begin{tabular}{@{}c@{}}$\psi$ \\ $[^\circ]$\end{tabular} & 
\begin{tabular}{@{}c@{}}$\zeta$ \\ $[\ln(\mathrm{kg}/\mathrm{m}^3)]$\end{tabular} & 
\begin{tabular}{@{}c@{}}$\sigma$ \\ $[^\circ]$\end{tabular} & 
\begin{tabular}{@{}c@{}}$\frac{L}{D}$ \\ $\text{[n.d.]}$\end{tabular} & 
\begin{tabular}{@{}c@{}}$\beta$ \\ $[\mathrm{kg}/\mathrm{m}^2]$\end{tabular} \\
    \hline
    Baseline & 1000 & 190.05 & -9.76 & 24.93 & -10.58 & 45 & -23.32 & 78 & 0.25 & 145 \\ 
    High Vel. & 1000 & 190.05 & -9.76 & 27.21 & -11.19 & 45 & -23.32 & 75.5 & 0.25 & 145 \\ 
    \hline
\end{tabular}
\end{table}

\begin{figure}[hbt]
	\centering
	\includegraphics{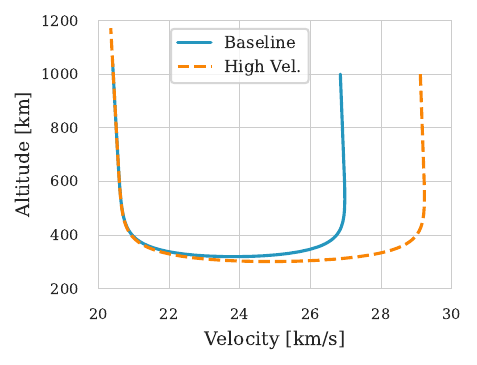}
  \caption{Nominal aerocapture trajectories.}
  \label{fig:example-traj}
\end{figure}

The nominal trajectories in altitude-velocity space are shown in \cref{fig:example-traj}.

\begin{figure}[hbt]
	\centering
	\includegraphics{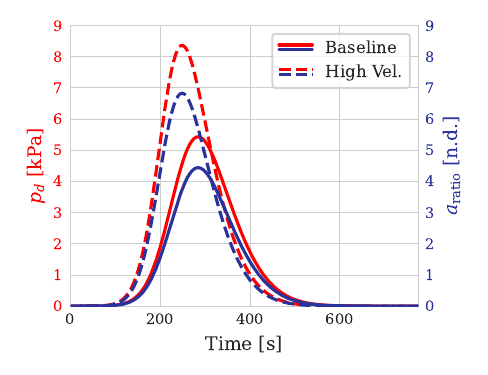}
  \caption{Dynamic pressure and acceleration ratio profiles for the nominal trajectories.}
  \label{fig:dynamic_pressure}
\end{figure}

The dynamic pressure profile, $p_d = \frac{1}{2} \rho V^2$, and ratio of gravitational to aerodynamic accelerations, $a_{\text{ratio}} = \sqrt{(L^2+D^2) / (g_r^2 + g_{\phi}^2)}$, for these trajectories are shown in \cref{fig:dynamic_pressure}. The high velocity scenario experiences a greater peak dynamic pressure due to increased deceleration during atmospheric flight. As high dynamic pressure and greater aerodynamic acceleration correspond to high nonlinearity in the dynamics, identifying this region is important in the following discussions.

Other key quantities of interest for aerocapture are the resulting apoapsis radius $r_a$, 
\begin{equation} \label{eqn:apoapsis_radius}
	r_a = a \left( 1 + \sqrt{1 - \frac{V^2 r^2 \cos^2 \gamma}{\mu a}} \right),
\end{equation}
where $a$ is the semi-major axis defined as:
\begin{equation}
	a = \frac{\mu}{\frac{2 \mu}{r} - V^2}, 
\end{equation}
and specific orbital energy $\varepsilon$,
\begin{equation} \label{eqn:specific_energy}
	\varepsilon = \frac{V^2}{2} - \frac{\mu}{r}.
\end{equation}

\section{Higher-Order Taylor Series for Perturbation Propagation} 
\label{sec:sttsdstts}

As seen in the previous section, the aerocapture problem is highly nonlinear, so linear propagation schemes are not typically sufficient to accurately propagate state perturbations. While small deviations may remain within the linear regime for the initial portion of flight, inherent nonlinearities can rapidly lead to inaccuracies when only using linear propagation. This work employs STTs to nonlinearly and efficiently propagate aerocapture state perturbations. Previous work has shown that accurate reduced-dimension DSTTs can be constructed for orbital scenarios; however, this approach has not been extended to atmospheric flight~\cite{boone_directional_2023,zhou_time-varying_2024,zhou_efficient_2025}. Once the STTs are obtained for a particular flight scenario, they contain rich information about the dynamics, which can be used to construct the reduced-dimension DSTTs that efficiently propagate perturbations. Understanding how to accurately model perturbations is the basis for uncertainty propagation, which will be the focus of future work.

\subsection{Review of State Transition Tensors}
\label{sec:STTs}

STTs are a higher-order extension of the commonly used state transition matrix (STM). The solution to  a dynamic system with state vector $\mathbf x \in \mathbb{R}^{n}$ can be represented using the solution flow, $\boldsymbol{\phi}$, where 
\begin{equation}
	\mathbf{x}(t) = \boldsymbol{\phi}(t; \mathbf{x}_y, t_y),
\end{equation}
for given initial condition $\mathbf{x}_y$ at time $t_y$~\cite{park_nonlinear_2006,park_nonlinear_2007}. The STTs are the partial matrices of this solution flow with respect to the state at a previous time $\mathbf{x}_{y}$,
\begin{equation} \label{eqn:stt_def}
	\phi_{(t_z, t_y)}^{i, \gamma_1\cdots\gamma_p} = \frac{\partial^p\boldsymbol{\phi}^i(t_z;\xvec_y, t_{y})}{\partial x_{y}^{\gamma_1} \cdots \partial x_{y}^{\gamma_p}}.
\end{equation}
This time indexing will be used throughout the paper for generality to any $\Delta t = t_z - t_y$. In addition, index notation is used throughout this paper, where superscripts indicate the components of a tensor and subscripts indicate the time of a vector or the times a matrix or tensor maps between. Repeated superscript indices are summed over, and commas in superscripts separate components that are summed over from those which are not. The number of superscripts indicates the order of the tensor. For example, a second-order tensor is a matrix, a third-order tensor is a three-dimensional tensor, and so on.\footnote{Note that the STM is the first-order STT which is a matrix, and the second-order STT is a third-order tensor. The order of an STT is determined by the number of copies of $\delta \mathbf{x}$ with which it is contracted.} 

The STTs can be obtained by integrating the variational equations, which can be found in prior work~\cite{park_nonlinear_2006}. A perturbation can be propagated through $m^{\text{th}}$-order STTs following~\cite{park_nonlinear_2006}:
\begin{equation} \label{eqn:stt_pert}
	\delta x_z^i = \sum_{p=1}^m \frac{1}{p!} \phi_{\left(t_z, t_y\right)}^{i, \gamma_1 \gamma_2 \ldots \gamma_p} \delta x_y^{\gamma_1} \delta x_y^{\gamma_2} \ldots \delta x_y^{\gamma_p}
\end{equation}

As with the STM, once a set of STTs at points across a time domain $[t_0,t_f]$ are obtained, they can be manipulated to map over another time domain within the original time domain. Once the STTs are obtained on the interval $[t_0,t_f]$, the STT mapping from time $t_y$ to $t_z$, where $t_y, t_z \in [t_0,t_f]$, can be computed algebraically~\cite{park_nonlinear_2007,cunningham_interpolated_2023}. This is especially relevant for spacecraft missions with minimal onboard storage: instead of storing a different set of STTs for each time regime of interest, a single set can be stored so long as it contains the desired intermediate times $t_y, t_z \in [t_0,t_f]$. While it may be useful to use STTs mapping from $t_k$ to $t_{k+1}$ for navigation and filtering purposes, guidance may desire to find a higher-order solution from $t_k$ to $t_f$. Uncertainty propagation methods could employ a mapping from $t_0$ to $t_k$. This property allows a perturbation to the state to be mapped nonlinearly onboard over any time domain with a single set of STTs computed offline. 

\subsection{State Transition Tensors for Aerocapture}
\label{sec:stt_example}

Studies investigating the accuracy of higher-order Taylor series methods for aerocapture uncertainty propagation remain limited~\cite{yu_sensitivity_2025,calkins2025dynamics,calkins_grace_e_efficient_2024}. Although STTs have demonstrated accurate performance for orbital uncertainty propagation~\cite{park_nonlinear_2006,park_nonlinear_2007,khatri_nonlinear_2023,armellin_asteroid_2010}, their applicability to highly nonlinear aerocapture dynamics may be limited if large perturbations reduce the validity of local Taylor series expansions. Here we present results demonstrating the accuracy of STTs for aerocapture perturbation propagation for the two scenarios described in \cref{sec:aerocapture}.

10,000 $\delta \xvec_0$ samples were taken from the Gaussian distribution in \cref{tab:init_cov} and propagated through the STM ($\mathbf{\Phi}$), second-order STTs (${\phi}^{[2]}$), and third-order STTs (${\phi}^{[3]}$).  The initial natural log of density, $\zeta$, was not dispersed as variations in initial velocity and flight path angle lead to dispersions in this parameter (see \cref{eqn:density}). These dispersions result in $\pm 50$\% variation from the nominal density within the samples. 

\begin{table}[h!]
\centering  
\caption{Initial Standard Deviations.}
\label{tab:init_cov}
\begin{tabular}{ccccccc}
    \hline
    3$\sigma_r$ [km] & 3$\sigma_\theta$ [$^\circ$] &  3$\sigma_\phi$ [$^\circ$] &  3$\sigma_V$ [m/s] & 3$\sigma_\gamma$ [$^\circ$] & 3$\sigma_\psi$ [$^\circ$] & 3$\sigma_{\zeta}$ [$\ln$(kg/m$^3$)] \\
    \hline
    $10$ & 0.1 & 0.1 & 5 & 0.1 & 0.1 & 0\\ \hline
\end{tabular}
\end{table} 

\begin{figure}[hbt]
	\centering
	\includegraphics{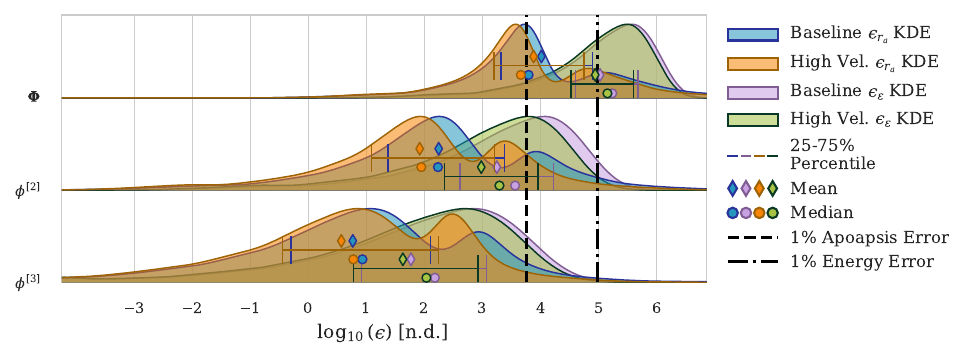}
  \caption{STT propagation accuracy for the baseline and high velocity aerocapture scenarios.}
  \label{fig:stt_prop}
\end{figure}

Each $\delta \xvec_0$ was propagated through a Taylor series expansion computed from $(t_f, t_0)$ using \cref{eqn:stt_pert}. The ridge plot in \cref{fig:stt_prop} compares the statistics of the standard logarithm of terminal apoapsis radius error and terminal energy error between the Taylor series propagated solution and the integrated solution for each $\delta \xvec_0$: 
\begin{equation}
	\log_{10} (\epsilon_{r_a}) = \log_{10} \left( \left| r_{a,\text{analytical}} - r_{a, \text{integrated}} \right| \right),
\end{equation}
for apoapsis radius in km, and
\begin{equation}
	\log_{10} (\epsilon_{\varepsilon}) = \log_{10} \left( \left| \varepsilon_{\text{analytical}} - \varepsilon_{\text{integrated}} \right| \right),
\end{equation}
for specific energy in m$^2$/s$^2$. The absolute value is required as terminal apoapsis is negative for hyperbolic trajectories and terminal energy is negative for captured trajectories. 

 To compute an empirical probability density function using kernel density estimation (KDE), the standard logarithm of terminal apoapsis radius and energy errors was downsampled to the 0--99$^\text{th}$ percentile of the data to reduce noise from outliers. The KDE of the data was then computed using \texttt{scipy.stats.gaussian\_kde} function from SciPy~\cite{2020SciPy-NMeth}. The KDE of the standard log of these errors\textemdash along with the 25--75$^\text{th}$ percentile, mean, and median of the data\textemdash are plotted in \cref{fig:stt_prop} for both the baseline and high velocity scenarios. 

The $x$-axis corresponds to the order of magnitude of the apoapsis radius error. For both scenarios, the nominal apoapsis radius is approximately $5.7\times 10^6$ km. An $x$ value of 2 for apoapsis corresponds to a 100 km error in apoapsis prediction, which is approximately a 0.0018\% miscalculation. Additionally, the nominal terminal energy is approximately $-9.7\times 10^6$ J/kg.  An $x$ value of 3 for energy corresponds to a 1,000 m$^2$/s$^2$ error, which is about a 0.010\% miscalculation. 1\% apoapsis and energy error lines are included in \cref{fig:stt_prop} for clarity. 

While the energy error results are unimodal, the apoapsis errors are multimodal. The secondary apoapsis error mode occurs at errors approximately one to two orders of magnitude larger than the dominant mode. These larger errors arise due to a singularity in the apoapsis expression (see \cref{eqn:apoapsis_radius}) between captured (positive terminal apoapsis radius) and hyperbolic (negative terminal apoapsis radius) cases. The high error mode corresponds primarily to hyperbolic or near-escape trajectories: near the capture boundary, small state errors can produce large apoapsis variations. While the STT solutions can predict escaped trajectories, despite the STT expansion point being a captured trajectory, the STTs are less accurate for these escaped trajectories. This phenomenon is not present in the energy errors, as the orbital energy varies continuously across the capture boundary. Energy is less sensitive to small errors in the state, transitioning smoothly from positive values for hyperbolic trajectories to negative values for captured trajectories.

As expected, increasing STT order increases propagation accuracy: the STM results in a mean apoapsis error of 10,000 km, the second-order STT results in a mean apoapsis error of 100 km, while the third-order STT results in a mean apoapsis error of less than 10 km. There is little difference in propagation accuracy between the baseline and high velocity scenarios, indicating the STTs are accurate for a range of aerocapture initial conditions. In addition, third-order STTs result in less than 1\% apoapsis and terminal energy misprediction, even for hyperbolic trajectories. Thus, these results demonstrate that STTs are accurate for aerocapture perturbation propagation for performance analysis. 
  
However, higher-order STTs require $\sum_{i=2}^{p} n^i$ terms to adequately capture the nonlinear effects, where $p$ is the STT order and $n$ is the number of states. It is computationally expensive to integrate the $\sum_{i=1}^{p+1} n^i$ variational equations for an $p^{\text{th}}$-order STT, and it can be complex to perform tensor contractions to propagate perturbations (or moments of a distribution for an uncertainty propagation application). For the 7-state aerocapture system, 399 equations must be integrated to construct the second-order STT, while 2,800 equations must be integrated to construct the third-order STT. Thus, producing a reduced-dimension representation of the STTs is advantageous.

\subsection{Review of Directional State Transition Tensors}

As discussed above, reducing the dimensionality of the STTs is beneficial. Multiple studies have found that many of the higher-order STT terms are negligible for orbital equations of motion~\cite{boone_directional_2023,vittaldev_multidirectional_2016}. The central challenge that follows is to construct a reduced-dimension STT that has a similar effect to the full STT when used for perturbation propagation. 

DSTTs are constructed through aligning the STTs with sensitive directions in the dynamics and neglecting the STT terms in the stable directions. Let $\mathbf{R}_{[p]} \in \mathbb{R}^{l\times n}$, where $l\leq n$, be a linear transformation matrix where $l$ is the number of sensitive directions and $p$ is the DSTT order, which is the same as the STT order it is constructed from. The rows of $\mathbf{R}_{[p]}$ are the orthogonal sensitive directions with which we seek to align the STTs. The DSTTs propagate the reduced-dimension perturbation $\delta \mathbf{y} = \mathbf{R}_{[p]} \delta \mathbf{x}$ through time instead of the full state $\delta \mathbf{x}$. The STT derivatives are taken with respect to the state vector $\mathbf{x} \in \mathbb{R}^n$, as seen in \cref{eqn:stt_def}, whereas the DSTTs can be seen as ``rotated and projected'' STTs such that their derivatives are taken with respect to another basis, $\mathbf{y} \in \mathbb{R}^l$. The second- and third-order DSTTs can be obtained  from the STTs by employing the chain rule, using the relations \cite{boone_directional_2023},
\begin{align}
\psi_{[2]}^{i, \gamma_1 \gamma_2}&=\phi^{i, \kappa_1 \kappa_2} R_{[2]}^{\gamma_1, \kappa_1} R_{[2]}^{\gamma_2, \kappa_2}, \qquad \text{and} \\
\psi_{[3]}^{i, \gamma_1 \gamma_2 \gamma_3}&=\phi^{i, \kappa_1 \kappa_2 \kappa_3} R_{[3]}^{\gamma_1, \kappa_1} R_{[3]}^{\gamma_2, \kappa_2} R_{[3]}^{\gamma_3, \kappa_3}, 
\end{align}
where $\phi^{i, \kappa_1 \ldots \kappa_p}$ are the elements of the $p^{\text{th}}$-order STTs and $\boldsymbol{\psi}_{[p]}$ is the $p^{\text{th}}$-order DSTT. Time indices are omitted for clarity. This expression differs from the original expression presented in~\cite{boone_directional_2023}, as it does not assume that the same $\mathbf{R}$ matrix is used for each DSTT order. Choosing $l < n$ reduces the number of terms required to nonlinearly propagate perturbations to the trajectory which significantly lowers storage and computation requirements. For example, a second-order STT has $n^3$ elements, but a second-order DSTT with $l=1$ has only $n$ terms. Thus, the DSTTs with $l < n$ are a reduced-dimension, higher-order representation of the solution flow. 

In order for DSTTs to be effective, the matrix $\mathbf{R}_{[p]}$ through which the STTs are transformed to reduce the basis dimension must be a direction which captures the higher-order effects of the STTs, and the dimension of the reduced basis must be sufficient to capture the higher-order effects of the dynamics. In previous studies, the rows of rotation direction matrix $\mathbf{R}_{[p]}$, for all orders $p$, have been chosen to be the eigenvectors of the largest $l$ eigenvalues of the second-order CGT~\cite{boone_directional_2023}, where the second-order CGT is defined as:
\begin{equation} \label{eqn:cgt}
	\boldsymbol{\mathcal{C}}^{[2]}_{(t_z, t_y)} = \boldsymbol{\Phi}^{\top}_{(t_z, t_y)}	\boldsymbol{\Phi}_{(t_z, t_y)},	
\end{equation}
where $\boldsymbol{\Phi}_{(t_z, t_y)}$ is the state transition matrix from $t_y$ to $t_z$. The rationale for constructing DSTTs using second-order CGT eigenpairs is discussed in \cref{sec:dynamics_analysis}. For a given propagation interval, the eigenvectors defining $\mathbf{R}_{[p]}$ are obtained by evaluating the CGT over the relevant regime. However, $\mathbf{R}_{[p]}$ must be recomputed for each reference trajectory and propagation interval, which becomes cumbersome when many trajectories or intervals are considered. To mitigate this issue, Zhou et al. propose a time-varying DSTT formulation that approximates the top $l$ eigenpairs of the second-order CGT over time~\cite{zhou_time-varying_2024}; however, this approach sacrifices accuracy and requires $l > 1$ to ensure the dominant stretching direction is captured.

Similarly to STTs, DSTTs can be used to propagate a perturbation to the state. In order to retain first-order accuracy, the full STM is used and all higher-order STTs are represented as DSTTs in the reduced basis. A deterministic perturbation can be propagated through the $m^{\text{th}}$-order DSTTs from $t_y$ to $t_z$ as \cite{boone_directional_2023}:
\begin{equation} \label{eqn:dstt_deterministic}
    \delta x_{z}^i \simeq \phi^{i, \kappa_1} \delta x_y^{\kappa_1}+\sum_{p=2}^m \frac{1}{p!} \psi_{[p]}^{i, \gamma_1 \gamma_2 \ldots \gamma_p} \delta y_y^{\gamma_1} \delta y_y^{\gamma_2} \ldots \delta y_y^{\gamma_p}.
\end{equation}
Time subscripts are omitted from the STM and DSTTs for clarity. The $\delta \mathbf{y}$ for order $p$ is computed using $\mathbf{R}_{[p]}$.  

For orbital scenarios, DSTTs constructed using the top $l$ magnitude eigenpairs of the second-order CGT have been sufficient ~\cite{boone_directional_2023,boodram_efficient_2022,zhou_efficient_2025,zhou_time-varying_2024}. However, some findings in this previous work do not hold true for the aerocapture. Although previous work found little variation between $\mathbf{R}_{[p]}$'s for each time domain across an orbit~\cite{boodram_efficient_2022},  each DSTT propagation interval ($t_z, t_y$) has a different $\mathbf{R}_{[p]}$ in this work because the direction changes greatly throughout the trajectory. While the second-order CGT eigenvectors proved sufficient to construct DSTTs that preserved the effects from the STTs for orbital scenarios~\cite{boone_directional_2023,boodram_efficient_2022,zhou_efficient_2025,zhou_time-varying_2024}, this method has been found to be inaccurate for aerocapture~\cite{calkins2025dynamics,calkins_grace_e_efficient_2024}. In addition, the direction of maximum stretching in the nonlinear dynamics was previously shown to be well-aligned with the direction of linear maximum stretching (the eigenvector of the maximum eigenvalue of the second-order CGT)~\cite{boone_directional_2023}, and this direction was relatively constant throughout the orbit~\cite{boodram_efficient_2022}. However, the analysis in the following sections will show that these findings do not hold true for aerocapture.  

\section{Aerocapture Dynamics Analysis Using Taylor Series}
\label{sec:dynamics_analysis}

Once STTs have been computed along the full trajectory, they enable a variety of dynamics analysis tools. The following section shows how STTs can be employed to determine the sources of temporal variability of the maximal \textit{linear} stretching direction. We will then use STTs to facilitate the identification of directions of maximum \textit{nonlinear} stretching for the entire state, subsets of the state, or nonlinear functions of the state. The information gained from these tools underscore a key benefit of STTs for perturbation propagation: once the STTs are obtained, a variety of information about the dynamics can be computed and used to construct more accurate reduced-order propagation methods. 

\subsection{Decomposed Cauchy Green Tensors}

Jenson et al. proved that the initial perturbation direction that maximizes the final state perturbation magnitude for the linear dynamics is the eigenvector corresponding to the maximum eigenvalue of the second-order CGT, computed using \cref{eqn:cgt}~\cite{jenson_bounding_2024}. A perturbation along this direction will yield the largest magnitude change in the final state. While previous studies have found that this direction is near constant for certain orbital mechanics problems~\cite{boone_directional_2023,boodram_efficient_2022}, the authors found that this was not the case for aerocapture~\cite{calkins_grace_e_efficient_2024,calkins2025dynamics}. To demonstrate the cause of this discrepancy, we separate the aerocapture dynamics into two functions: one for the conservative rotational and gravitational accelerations, and one for the dissipative aerodynamic accelerations. The CGTs of each of these dynamics are evaluated separately to demonstrate the cause of temporal variation in the maximal second-order CGT eigenvector direction.

Given a vector-valued nonlinear dynamics function $\dot{\xvec} = f(\mathbf{x})$,
\begin{equation} \label{eqn:fsum}
    f(\xvec) = f_C(\xvec) + f_D(\xvec),
\end{equation}
where $f_C(\xvec)$ are the conservative dynamics and $f_D({\xvec})$ are the dissipative dynamics, the first-order Taylor series expansion of $f$ about the reference trajectory $\bar{\xvec}$ is:
\begin{equation} \label{eqn:first_order_taylor}
    f(\xvec) \approx f(\bar{\xvec}) + \left. \frac{\partial f}{\partial \xvec} \right|_{\bar{\xvec}} \delta x = f(\bar{\xvec}) + \mathbf{A} \delta \xvec, 
\end{equation}
where $\mathbf{A}=\left. \frac{\partial f}{\partial \xvec} \right|_{\bar{\xvec}}$. Evaluating $f(\xvec)$ with \cref{eqn:fsum} at the reference state $\bar{\xvec}$ ,
\begin{equation}
    f(\bar{\xvec}) = f_C(\bar{\xvec}) + f_D(\bar{\xvec}),
\end{equation}
and taking its first derivative, 
\begin{equation}
    \left. \frac{\partial f}{\partial \xvec} \right|_{\bar{\xvec}} = \left. \frac{\partial f_C}{\partial \xvec} \right|_{\bar{\xvec}} + \left. \frac{\partial f_D}{\partial \xvec} \right|_{\bar{\xvec}} = \mathbf{A}_C + \mathbf{A}_D,
\end{equation}
the first-order Taylor series expansion in \cref{eqn:first_order_taylor} becomes:
\begin{equation}
    f(\xvec) \approx f_C(\bar{\xvec}) + f_D(\bar{\xvec}) + (\mathbf{A}_C + \mathbf{A}_D) \delta \xvec,
\end{equation}

A STM solves the variational equation:
\begin{equation}
    \dot{\boldsymbol{\Phi}} = \mathbf{A} \boldsymbol{\Phi}.
\end{equation}
STMs for the decomposed dynamics can be found using the separated variational equations:
\begin{equation}
    \dot{\boldsymbol{\Phi}}_C = \mathbf{A}_C \boldsymbol{\Phi}_C, \quad \text{and} \quad \dot{\boldsymbol{\Phi}}_D = \mathbf{A}_D \boldsymbol{\Phi}_D.
\end{equation}
The directions that maximize final perturbation magnitude under the $f_i$ linearized dynamics function can be shown to be the direction corresponding to the maximal eigenvector of the CGT computed using $\boldsymbol{\Phi}_i$~\cite{jenson_bounding_2024,jenson_semianalytical_2023}.
\footnote{Note that the STMs of the decomposed dynamics cannot be superposed to reconstruct the STM for the dynamics of the full system, because the STMs are functions of the underlying nonlinear dynamics.} 

\begin{figure}[htb]
     \centering
     \includegraphics{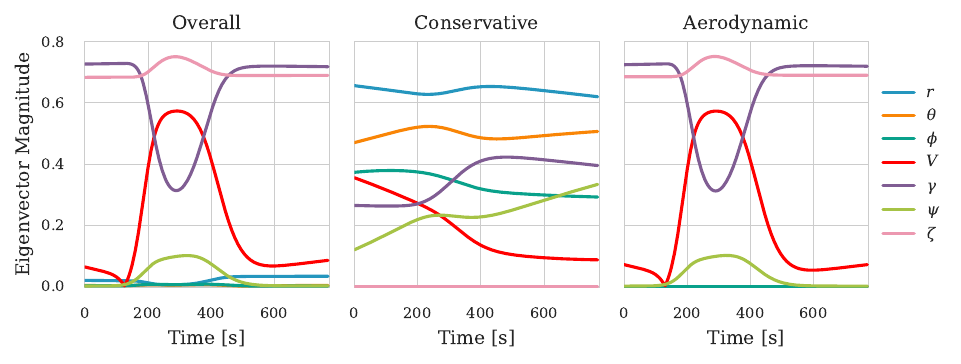}
     \caption{Eigenvectors associated with the maximum eigenvalue for each decomposed dynamics second-order CGT from $(t_{k+1}, t_k)$ for the baseline scenario.}
     \label{fig:separated_cgts}
\end{figure}

The CGTs using the decomposed STM were evaluated along the baseline scenario reference trajectory on the interval $(t_{k+1}, t_k)$, where $\Delta t = t_{k+1} - t_k = 5$ seconds, and the eigenvector corresponding to the maximum eigenvalue is shown in \cref{fig:separated_cgts}. The overall direction is clearly time-varying, and the large magnitude changes in the velocity and density coordinates result from the aerodynamic portion of the equations of motion, while the conservative equations of motion have a relatively constant direction throughout the entire trajectory. As expected, density plays no role in the conservative dynamics. The region of high variation in eigendirection corresponds to the ratio of aerodynamic to gravitational accelerations exceeding one in \cref{fig:dynamic_pressure}. This finding demonstrates why a time-varying direction is required to construct DSTTs for aerocapture, even though this is not essential for orbital scenarios.

\begin{figure}[hbt]
	\centering
  \includegraphics{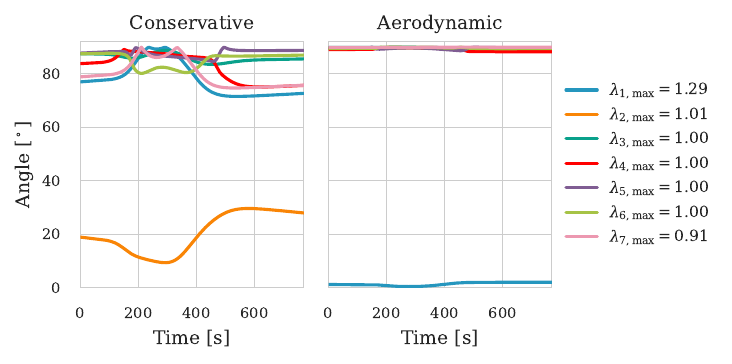}
  \caption{Angle between dominant decomposed dynamics CGT eigenvector and all overall CGT eigenvectors for the baseline scenario.}
  \label{fig:overall_modes}
\end{figure}

\cref{fig:separated_cgts} also shows that the maximum eigenvector overall is more similar to the aerodynamic eigenvector, indicating that the aerodynamic accelerations dominate the equations of motion. As the magnitude of a CGT eigenvalue reflects the local stretching or contraction rate, eigenvectors with larger eigenvalues denote directions along which the perturbation are most amplified in the dynamics. \cref{fig:overall_modes} shows the angle between the eigenvector corresponding to each mode of the overall CGT and the maximal eigenvectors of the decomposed aerodynamic and conservative CGTs. The maximum value of the eigenvalue of the overall dynamics corresponding to each eigenvector is provided in the legend. The  dominant CGT eigenvector of the aerodynamic CGT clearly corresponds to the dominant mode of the overall CGT, as the angle is near zero for all times between the dominant aerodynamic eigenvector and dominant overall eigenvector (the eigenvector of maximum eigenvalue $\lambda_1$ in the overall dynamics). The dominant mode of the conservative CGT best matches the second mode of the overall dynamics, corresponding to eigenvalue $\lambda_2$, which has a 22\% difference in maximum magnitude from $\lambda_1$. 

\subsection{Tensor Eigenpairs of Higher-Order Cauchy Green Tensors}
\label{sec:hocgts}

Aerocapture is a highly nonlinear problem, and the second-order CGT only considers stretching due to the linear dynamics. While previous studies found that a particularly sensitive direction in the second-order CGT indicated that the higher-order terms along this direction are significantly larger than the other terms~\cite{boone_directional_2023}, this is not true for aerocapture. To find a dynamics-informed direction that considers nonlinearities, the tensor eigenpairs of the HOCGTs can be evaluated, and these directions can be used to directionalize the DSTTs \cite{jenson_bounding_2024,jenson_semianalytical_2023}. 

To find the direction of initial perturbation that results in the largest magnitude final state perturbation, the squared $L^2$ norm of $\mathbf{x}$ at a later time with respect to the state at a prior time can be evaluated \cite{jenson_bounding_2024}. When using a linear propagation of the state, this expression leads to the definition of the second-order CGT:
\begin{equation} \label{eqn:CGT}
    \left|\left|\boldsymbol{\Phi}_{(t_z, t_y)} \xvec_y \right|\right|^2 = \xvec_y^{\top} \boldsymbol{\Phi}_{(t_z, t_y)}^{\top}  \boldsymbol{\Phi}_{(t_z, t_y)} \xvec_y = \xvec_y^{\top}  \boldsymbol{\mathcal{C}}^{[2]}_{(t_z, t_y)} \xvec_y.
\end{equation}

When considering higher-order terms of the Taylor series expansion in the solution for $\xvec_z$, the squared norm of the state at $t_z$ with respect to the state at $t_y$ is:
\begin{equation} \label{eqn:HOCGTs}
    \xvec_z^{\top} \xvec_z = \boldsymbol{\mathcal{C}}^{[2]} \xvec_y^2 + \tilde{\boldsymbol{\Phi}}^{[3]} \xvec_y^3 + \tilde{\boldsymbol{\Phi}}^{[4]} \xvec_y^4 + \ldots,
\end{equation}
and the higher-order tensor maps for third- and fourth-order are given by:
\begin{align}
\tilde{\Phi}^{[3], \kappa_1 \kappa_2 \kappa_3}&=\phi^{i, \kappa_1} \phi^{i, \kappa_2 \kappa_3}, \quad \text{and} \\
\tilde{\Phi}^{[4], \kappa_1 \kappa_2 \kappa_3 \kappa_4}&=\frac{1}{3} \phi^{i, \kappa_1} \phi^{i, \kappa_2 \kappa_3 \kappa_4}+\frac{1}{4} \phi^{i, \kappa_1 \kappa_2} \phi^{i, \kappa_3 \kappa_4}.
\end{align}

Jenson and Scheeres showed that a perturbation in the direction of the eigenvector corresponding to the maximum eigenvalue of the HOCGTs will cause the maximum $L^2$ norm magnitude perturbation in the final state~\cite{jenson_bounding_2024}. Thus, while the second-order CGT eigenpair can be use to determine the direction of maximum stretching due to the \textit{linear} dynamics, the HOCGT eigenpairs can be used to determine the direction of maximum stretching due to the \textit{higher-order} dynamics. The higher-order tensor z-eigenpair problem is similar to the matrix eigenpair problem, $\mathbf{A}\mathbf{v} = \lambda \mathbf{v}$, but extended to higher orders, $\mathcal{T}^{[m]} \mathbf{v}^{m-1} = \lambda \mathbf{v}$ for $\mathbf{v}^{\top}\mathbf{v} = 1$. These eigenpairs are found using the Shifted-Symmetric Higher-Order Power Method (SS-HOPM) algorithm~\cite{kolda_shifted_2011}. 

A few considerations were taken when utilizing SS-HOPM. As SS-HOPM requires symmetric tensors, we compute symmetric tensors $\boldsymbol{\mathcal{C}}^{[m]}$ such that $\tilde{\boldsymbol{\Phi}}^{[m]} = \boldsymbol{\mathcal{C}}^{[m]} \xvec^m$~\cite{jenson_bounding_2024} using the MATLAB Tensor Toolbox~\cite{bader_algorithm_2006}. These $\boldsymbol{\mathcal{C}}^{[m]}$ for $m > 2$ are the HOCGTs. Additionally, SS-HOPM does not guarantee convergence to the maximum eigenvalue, merely an eigenvalue, so 100 initial guesses are sampled from the unit ball and evaluated with SS-HOPM to find candidates $\mathbf{v}_i$. The vectors are de-duplicated by removing all vectors which have a dot product with another $\mathbf{v}_i$ greater than $\cos(10^{-3} \; \text{rad})$. The eigenvector corresponding to the largest eigenvalue of the remaining vectors is selected as the maximal direction. While this method does not guarantee convergence to the largest eigenvalue, especially when the eigenvalues are close in magnitude, it does improve convergence over using a single initial guess.

\subsection{Augmented Higher-Order Cauchy Green Tensors}

The HOCGTs described above find the direction that maximizes the final perturbation magnitude for the entire states. However, for certain guidance and navigation applications, we are interested in looking at the direction of sensitivity for key nonlinear states. In the following sections, we propose two augmented HOCGT formulations. First, we introduce selective HOCGTs (sCGTs), which can be used to find the direction that leads to the largest magnitude change for specific states which are known to be nonlinear. Second, we introduce quantity-of-interest HOCGTs (qCGTs), which find the direction that leads to the largest magnitude change for a quantity of interest that is a function of the state. 

\subsubsection{Selective Higher-Order Cauchy Green Tensors}
\label{sec:scgts}

While the HOCGTs can accurately identify directions of nonlinearity for the full state~\cite{jenson_bounding_2024}, certain states are more pertinent for performance analysis than others. For example, Muralidharan and Howell previously investigated the stretching directions from subsets of the solution flow using second-order CGTs for Near Rectilinear Halo Orbits~\cite{muralidharan_leveraging_2022,muralidharan_stretching_2023}. In particular, they found that stretching magnitudes computed from subsets including velocity perturbations at the final time were significantly larger than those computed considering the entire state or position coordinates. By designing stationkeeping maneuvers with minimal components along these stretching directions, the resulting trajectory had lower orbit maintenance costs. Instead of considering only the linear dynamics and changes in only position or velocity, we develop tensors which can be used to find the maximum stretching directions for arbitrary combinations of the state considering higher-order dynamics.

To understand dynamic sensitivity in these key states, we are most interested in obtaining the directions that correspond to the maximum magnitude change in those states, $\boldsymbol{\xi} = \mathbf{S} \mathbf{x} \in \mathbb{R}^{w}$, instead of the entire state vector, $\xvec$. $\mathbf{S} \in \mathbb{R}^{w\times n}$ is a constant selection matrix where $w \leq n$, where each row is a Cartesian basis vector along the state selected, and the selected states $\boldsymbol{\xi}$ are in $\mathbb{R}^w$. Here we derive sCGTs, whose tensor eigenpairs can be used to find the maximum stretching direction for the selected states.

The expression to propagate the selected states $\boldsymbol{\xi}$ from $t_y$ to $t_z$ can be written using STTs as: 
\begin{equation}
    \xi^i_z = \sum_{p=1}^m \frac{1}{p!} S^{i,j} \phi^{j, \kappa_1 \ldots \kappa_p}_{(t_z, t_y)} x_y^{\kappa_1} \ldots x_y^{\kappa_p},
\end{equation}
and the analog to HOCGT \cref{eqn:HOCGTs} in for sCGTs is:
\begin{equation} \label{eqn:scgt_def1}
    \boldsymbol{\xi}_z^{\top} \boldsymbol{\xi}_z = \boldsymbol{\mathcal{S}}^{[2]} \mathbf{x}_y^2 + \tilde{\boldsymbol{\mathcal{S}}}^{[3]} \mathbf{x}_y^3 + \tilde{\boldsymbol{\mathcal{S}}}^{[4]} \mathbf{x}_y^4 + \ldots,
\end{equation}
where $\boldsymbol{\mathcal{S}}^{[2]}$ is the second-order sCGT, which is already symmetric by definition. The third- and fourth-order $\tilde{\boldsymbol{\mathcal{S}}}^{[m]}$ coefficients are not supersymmetric in general. The coefficients are defined as:
\begin{align}
    \mathcal{S}^{[2], \kappa_1 \kappa_2} &= S^{i,j} \phi^{j, \kappa_1} S^{i,k} \phi^{k, \kappa_2}, \\
    \tilde{\mathcal{S}}^{[3], \kappa_1\kappa_2\kappa_3} &= S^{i,j} \phi^{j, \kappa_1} S^{i,j} \phi^{j, \kappa_2\kappa_3}, \quad \text{and} \\
    \tilde{\mathcal{S}}^{[4], \kappa_1\kappa_2\kappa_3\kappa_4} &= \frac{1}{3} S^{i,j} \phi^{j, \kappa_1} S^{i,j} \phi^{j, \kappa_2 \kappa_3 \kappa_4} + \frac{1}{4} S^{i,j} \phi^{j, \kappa_1 \kappa_2} S^{i,j} \phi^{j, \kappa_3 \kappa_4}.
\end{align}
These again must be symmetrized in order to solve the eigenproblem using SS-HOPM such that $\tilde{\boldsymbol{\mathcal{S}}}^{[m]} = {\boldsymbol{\mathcal{S}}}^{[m]} \xvec^m$. The ${\boldsymbol{\mathcal{S}}}^{[m]}$ are the higher-order sCGTs.

\subsubsection{Quantity-of-Interest Higher-Order Cauchy Green Tensors}
\label{sec:qcgts}

Similarly, certain stretching directions are more significant to key quantities of interest. For aerocapture, quantities of interest could include specific orbital energy (\cref{eqn:specific_energy}) or apoapsis radius (\cref{eqn:apoapsis_radius}), both of which are key factors for a successful aerocapture. If we primarily want to predict apoapsis radius precisely, we must identify what directions impact error for this direction most. Given a generic vector-valued quantity of interest function $\mathbf{q} = q(\mathbf{x}_z)$, qCGTs are defined to find a the direction which this quantity is most sensitive to. The Taylor series expansion of $\mathbf{q}$ about the state at previous time $t_y$, $\mathbf{x}_y$ is:
\begin{equation} \label{eqn:q_taylor_series}
    \begin{split}
     q_z^i \simeq \sum_{p=1}^M \frac{1}{p!} \frac{\partial^p q_z^i}{\partial x_y^{\gamma_1} \ldots \partial x_y^{\gamma_p}} x_y^{\gamma_1} \ldots x_y^{\gamma_p}.
    \end{split}
\end{equation}

To derive the qCGTs, we are interested in coefficients of the quantity $\mathbf{q}^{\top} \mathbf{q}$, namely:
\begin{equation}
	    \mathbf{q}_z^{\top} \mathbf{q}_z = \boldsymbol{\mathcal{Q}}^{[2]} \mathbf{x}_y^2 + \tilde{\boldsymbol{\mathcal{Q}}}^{[3]} \mathbf{x}_y^3 + \tilde{\boldsymbol{\mathcal{Q}}}^{[4]} \mathbf{x}_y^4 + \ldots
\end{equation} 
Because $\mathbf{q}$ is a function of the the state at $t_z$, but the state perturbation is propagated from $t_y$, the chain rule is employed to relate the change in the later state to the change in the prior state. The first-order coefficient in the expansion in \cref{eqn:q_taylor_series} is:
\begin{equation}
    \frac{\partial q_z^i}{\partial x_y^{\gamma_1}} = \frac{\partial q_z^i}{\partial x_z^{\gamma_2}} \frac{\partial x_z^{\gamma_2}}{\partial x_y^{\gamma_1}}.
\end{equation}
By noting that 
\begin{equation}
    \frac{\partial x_z^{\gamma_2}}{\partial x_y^{\gamma_1}} = \phi^{\gamma_2,\gamma_1}_{(t_z, t_y)},
\end{equation}
and defining 
\begin{align}
    \eta^{i,\gamma_2} &=\frac{\partial q_z^i}{\partial x_z^{\gamma_2}}, \\
    \eta^{i,\gamma_2 \gamma_3} &=\frac{\partial^2 q_z^i}{\partial x_z^{\gamma_2} \partial x_z^{\gamma_3}}, \quad \text{and} \\ 
    \eta^{i,\gamma_2 \gamma_3 \gamma_4} &=\frac{\partial^3 q_z^i}{\partial x_z^{\gamma_2} \partial x_z^{\gamma_3} \partial x_z^{\gamma_4}},  
\end{align}
we can write the higher-order qCGT coefficients as:
\begin{align} \label{eqn:qCGT2} 
    \tilde{{\mathcal{Q}}}^{[2], \kappa_1 \kappa_2} &= D_{[1]}^{i,\kappa_1} D_{[1]}^{i,\kappa_2},\\
    \label{eqn:qCGT3} 
    \tilde{{\mathcal{Q}}}^{[3],\kappa_1 \kappa_2 \kappa_3} &= D_{[1]}^{i,\kappa_1} D_{[2]}^{i,\kappa_2 \kappa_3}, \quad \text{and} \\
	\tilde{{\mathcal{Q}}}^{[4], \kappa_1 \kappa_2 \kappa_3 \kappa_4} & = \frac{1}{3} D_{[1]}^{i,\kappa_1} D_{[3]}^{i,\kappa_2\kappa_3\kappa_4} + \frac{1}{4} D_{[2]}^{i,\kappa_1\kappa_2} D_{[2]}^{i,\kappa_4\lambda_4}.
\end{align}
where $\mathbf{D}_{[i]}$ is the $i^{\text{th}}$-order coefficient of the Taylor series expansion in \cref{eqn:q_taylor_series}, 
\begin{align}
	D_{[1]}^{i,\kappa_1} &= \eta^{i,j} \phi^{j, \kappa_1}, \\ 
	D_{[2]}^{i, \kappa_1 \kappa_2} &= \eta^{i,jl} \phi^{l,\kappa_2} \phi^{j,\kappa_1} + \eta^{i,j} \phi^{j,\kappa_1 \kappa_2}, \quad \text{and}\\ 
	D_{[3]}^{i, \kappa_1 \kappa_2 \kappa_3} & = \eta^{i,jlm} \phi^{m,\kappa_3} \phi^{l,\kappa_2} \phi^{j,\kappa_1} + \eta^{i,jl} \phi^{l,\kappa_2 \kappa_3} \phi^{j,\kappa_1} \\ \nonumber
	& + \eta^{i,jl} \phi^{l,\kappa_2} \phi^{j,\kappa_1 \kappa_3} + \eta^{i,jl}  \phi^{l,\kappa_3} \phi^{j,\kappa_1 \kappa_2} + \eta^{i,j} \phi^{j,\kappa_1 \kappa_2 \kappa_3}, 
\end{align}
Once the tensors are symmetrized such that $\tilde{\boldsymbol{\mathcal{Q}}}^{[m]} = \boldsymbol{\mathcal{Q}}^{[m]} \xvec^m$, the $\boldsymbol{\mathcal{Q}}^{[m]}$ are the higher-order qCGTs.

\subsection{Comparison of Higher-Order Cauchy Green Tensor Eigenpairs}
\label{sec:hocgt_comp}
We now compare the tensor eigenpairs of the novel augmented HOCGTs with the original HOCGT eigenpairs to demonstrate that (1) the eigenpairs vary for each HOCGT type and order, (2) the maximal mode in the higher-order dynamics changes throughout the trajectory, while the linear dynamics do not exhibit this modal variation, (3) the maximal stretching direction varies for small propagation intervals but stabilizes as $\Delta t = t_z - t_y$ increases, and (4) a perturbation in the direction of the eigenvector of the maximum eigenvalue of an augmented HOCGT results in the largest magnitude change in its objective.

\begin{figure}[htb]
     \centering
     \includegraphics{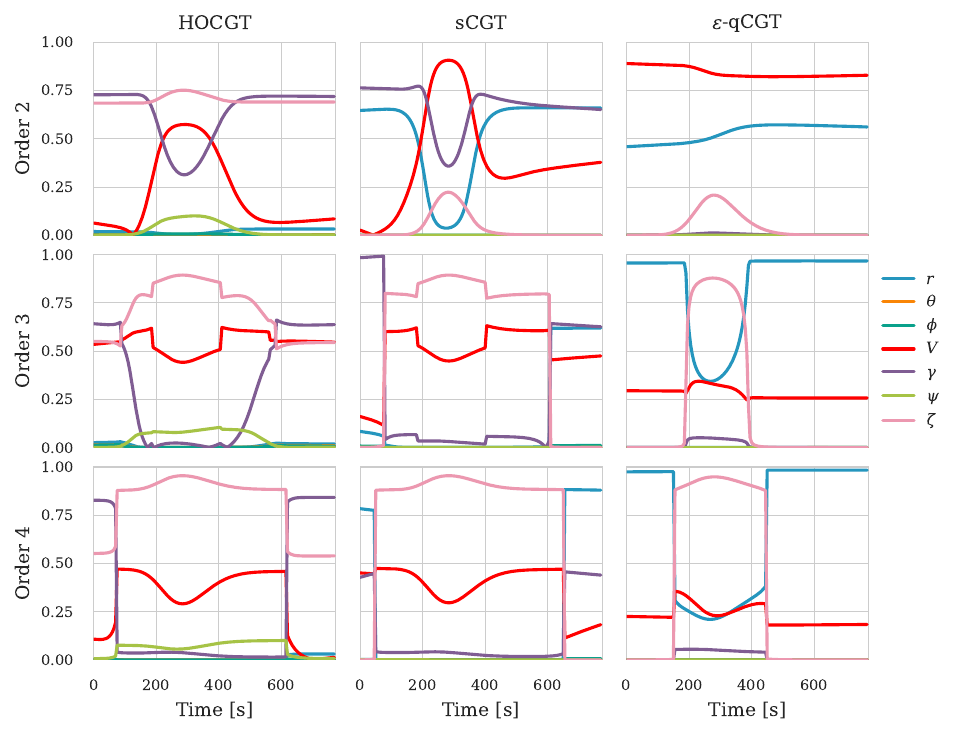}
     \caption{Eigenvectors of the maximum eigenvalue for different HOCGT constructions from $(t_{k+1}, t_k)$ for the baseline scenario. The sCGTs select position, velocity, and flight path angle; the $\varepsilon$-qCGTs use specific energy as the quantity-of-interest function.}
     \label{fig:hocgt_directions}
\end{figure}

\cref{fig:hocgt_directions} compares tensor eigenpairs corresponding to the maximum eigenvalue for the second- through fourth-order HOCGTs and augmented HOCGTs for $(t_{k+1}, t_k)$, where $\Delta t = t_{k+1} - t_k = 5$ seconds for the baseline scenario. The second-order eigenpairs are found using MATLAB's \texttt{eig} function, and the higher-order eigenpairs are found using the SS-HOPM algorithm from the Tensor Toolbox \cite{bader_tensor_2025}. The HOCGTs (see \cref{sec:hocgts}) are compared with the sCGTs for position, velocity, and flight path angle (see \cref{sec:scgts}), and qCGTs for specific energy using quantity of interest function \cref{eqn:specific_energy} (see \cref{sec:qcgts}). 

First, we see that the dominant direction varies significantly throughout the trajectory for this time regime. This is in direct contrast with previous work for orbital scenarios for propagation on $(t_{k+1}, t_k)$, which found that the unstable direction was rather constant throughout an orbit~\cite{boodram_efficient_2022}.  

When comparing orders, we can clearly see that the second-order stretching direction, which corresponds to linear dynamics, does not align with the higher-order directions for any HOCGT method. Therefore, using the second-order CGT direction is not appropriate for directionalizing second- and third-order DSTTs for aerocapture because it does not capture the direction of maximum stretching in the higher-order dynamics. Additionally, the sCGTs and $\varepsilon$-qCGTs do not have the same dominant eigenvector as the original HOCGTs. If the performance analysis goal is to track these key quantities, using the HOCGT directions for DSTT directionalization could lose higher-order information related to the coordinates important to the quantity of interest while keeping other less important directions.

\begin{figure}[hbt]
	\centering
	\includegraphics{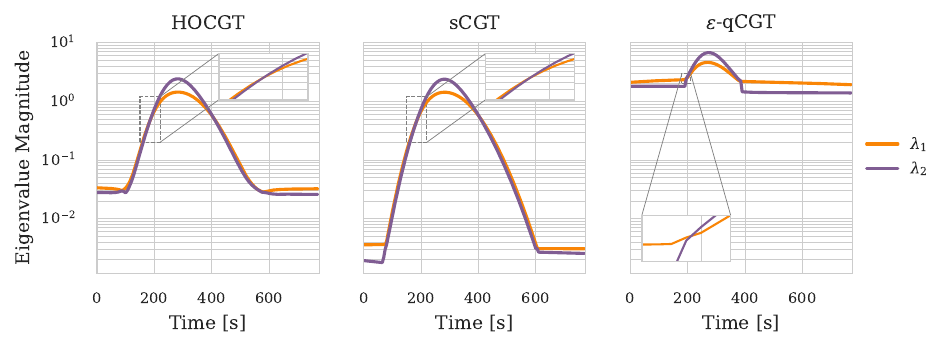}
  \caption{Top two eigenvalues of the third-order HOCGTs found from SS-HOPM for $(t_{k+1}, t_k)$ for the baseline scenario.}
  \label{fig:eval_diff}
\end{figure}

\cref{fig:hocgt_directions} also demonstrates that the dominant eigenpair for the higher-order dynamics switches throughout flight. The maximum eigenpair of the third-order HOCGTs and sCGTs changes at approximately 200 and 400 seconds, which causes the noticeable discontinuity in the eigenvector direction for $V$ and $\ln(\rho)$. The two largest eigenvalues, denoted $\lambda_1$ and $\lambda_2$, found from SS-HOPM for the third-order HOCGTs or augmented HOCGTs are shown in \cref{fig:eval_diff}. The relative magnitude of eigenvalues $\lambda_1$ and $\lambda_2$ switch throughout time. This indicates that the dominant mode of the nonlinear dynamics switches during maximum dynamic pressure, where the aerodynamic accelerations exceed the gravitational accelerations (as shown in \cref{fig:dynamic_pressure}), as gravitational accelerations become less significant and atmospheric accelerations dominate the dynamics. While the results on the interval $(t_{k+1}, t_k)$ are only presented for the baseline scenario, the conclusions are similar for the high velocity scenario. 

\begin{figure}[hbt]
	\centering
  \includegraphics{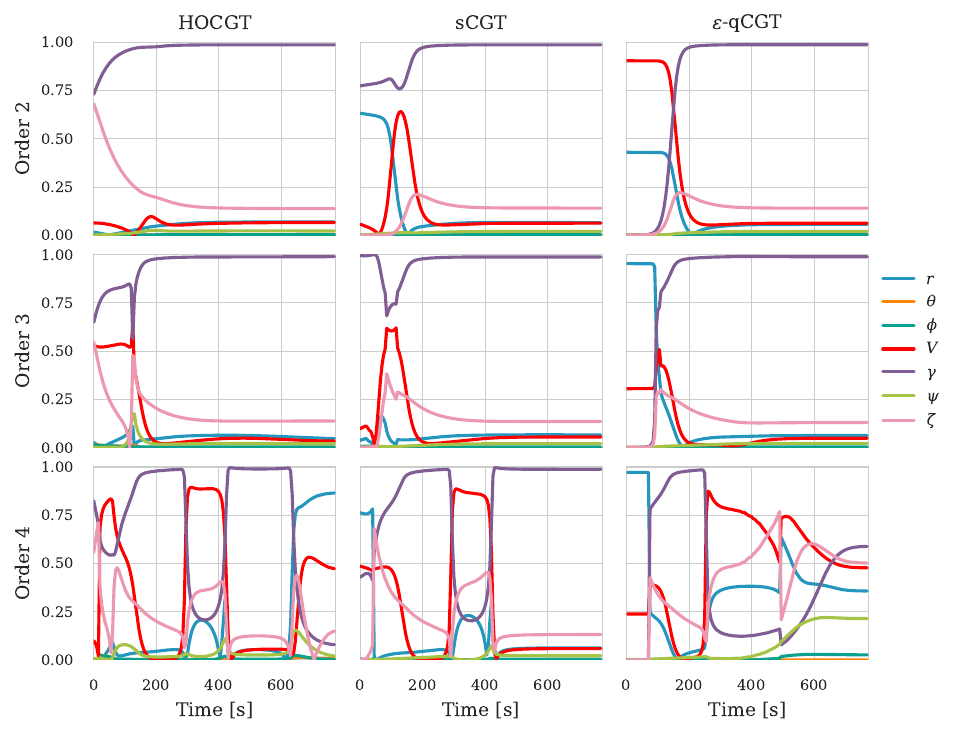}
  \caption{ Eigenvectors of the maximum eigenvalue of varying HOCGT orders from $(t_{k}, t_0)$ for the high velocity scenario.}
  \label{fig:hocgt_t0tk}
\end{figure}

While the dominant eigendirection varies significantly in time for the short propagation interval discussed above, as $\Delta t$ increases, the maximal stretching direction stabilizes to a similar vector for the second- and third-order HOCGTs and augmented HOCGTs. This can be seen in \cref{fig:hocgt_t0tk}, which depicts the maximal eigenvector for the interval $(t_k, t_0)$ for the high velocity scenario. While the direction varies over time and between orders for approximately the first 200 seconds of flight, as $t_k$ increases, the direction stabilizes. However, the direction does not stabilize for the fourth-order HOCGTs and augmented HOCGTs. This variation is due to mode switching, as discussed above. Thus, we cannot conclude that the maximal stretching direction stabilizes for all orders and HOCGT types for the aerocapture dynamics, despite this being the case for some dynamics and initial conditions. For example, for the baseline scenario on the interval $(t_k, t_0)$, the fourth-order maximal eigenvector of the HOCGT and sCGT does stabilize over time, while the fourth-order maximal eigenvector of the $\varepsilon$-qCGT varies temporally.

In addition to comparing the directions for different HOCGT constructions, we also verify that each HOCGT correctly identifies the direction that produces the largest final perturbation in the relevant objective (the total state norm for HOCGTs, selected states norm for sCGTs, and apoapsis radius for the $r_a$-qCGTs). To evaluate this, perturbations are propagated along seven orthogonal directions: the maximal third-order HOCGT or augmented HOCGT direction, $\mathbf{R}_{[3],(t_f,t_0)}$, and six orthogonal directions (which span the state space). These orthogonal initial perturbation directions are found using QR decomposition. The direction vector $\mathbf{R}_{[3],(t_f,t_0)}$ is inserted as the first column of the matrix:
\begin{equation}
	\mathbf{M} = \begin{bmatrix}
		\mathbf{R}_{[3],(t_f,t_0)} & \mathbf{e}_2 & \cdots & \mathbf{e}_7
	\end{bmatrix}, 
\end{equation}
where $\mathbf{e}_i$ are the standard basis vectors. A QR decomposition is applied, $\mathbf{M} = \mathbf{Q}\mathbf{R}_{\text{QR}}$, to produce orthonormal matrix $\mathbf{Q}$. Each column of the $\mathbf{Q}$ matrix can be extracted as 
\begin{equation}
	\mathbf{Q} = \begin{bmatrix}
		\mathbf{R}_{[3],(t_f,t_0)} & \mathbf{d}_2 & \cdots & \mathbf{d}_7
	\end{bmatrix}, 
\end{equation}
such that the first column of $\mathbf{Q}$ aligns with the direction vector $\mathbf{R}_{[3],(t_f,t_0)}$, while the remaining columns span its orthogonal complement. The remaining columns of the $\mathbf{Q}$, $\mathbf{d}_2 \ldots \mathbf{d}_7$, can be extracted to yield six perturbation directions which are mutually orthogonal to $\mathbf{R}_{[3],(t_f,t_0)}$. Each perturbation vector is scaled to a prescribed nondimensional magnitude of $10^{-3}$ to remain within the region of convergence of the STTs. 

\begin{figure}[hbt]
	\centering
	\begin{subfigure}[b]{0.33\textwidth}
         \centering
         \includegraphics[width=\textwidth]{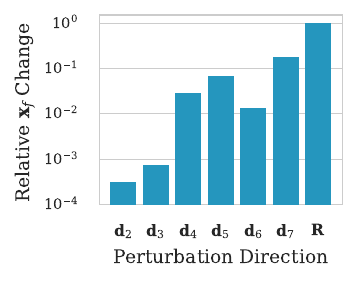}
         \caption{HOCGT}
         \label{fig:hocgt_mag}
     \end{subfigure}
     \hfill
     \begin{subfigure}[b]{0.33\textwidth}
         \centering
         \includegraphics[width=\textwidth]{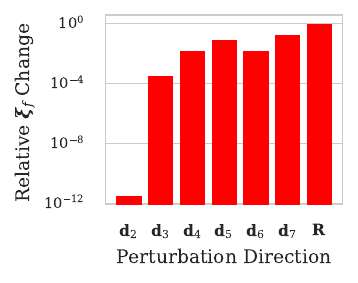}
         \caption{sCGT}
         \label{fig:scgt_mag}
     \end{subfigure}
     \hfill
     \begin{subfigure}[b]{0.33\textwidth}
         \centering
         \includegraphics[width=\textwidth]{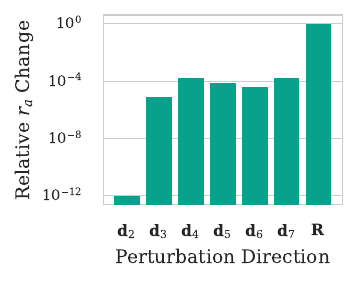}
         \caption{$r_a$-qCGT}
         \label{fig:qcgt_mag}
     \end{subfigure}
     \caption{Final state and quantity of interest change for orthogonal initial perturbation directions for the baseline scenario.}
     \label{fig:xf_magnitude}
\end{figure}

The perturbations are mapped through the dynamics over $(t_f, t_0)$ using numerical integration for the baseline scenario. \cref{fig:xf_magnitude} shows the relative final objective magnitudes for each direction set. We express each objective as a relative magnitude, normalized by the magnitude obtained from the maximal stretching direction, so we can interpret the result as a fraction for comparison. For the HOCGT maximal stretching direction in \cref{fig:hocgt_mag}, the objective is the norm of difference between the perturbed final state and the nominal final state for all seven directions, normalized by this quantity for the perturbation along the maximal stretching direction:
\begin{equation}
	\frac{\|\mathbf{x}_f^{\text{pert}} - \mathbf{x}_f^{\text{nom}}\|}{\|\mathbf{x}_f^{\text{pert}} - \mathbf{x}_f^{\text{nom}}\|_{\mathbf{R}_{[3],(t_f,t_0)}}}
\end{equation}
Likewise, the norm of the difference in the final perturbed selected states and the final nominal selected states normalized by this quantity for the perturbation along the maximal stretching direction is shown in \cref{fig:scgt_mag}:
 \begin{equation}
 	\frac{ \|\boldsymbol{\xi}_f^{\text{pert}} - \boldsymbol{\xi}_f^{\text{nom}}\|}{\|\boldsymbol{\xi}_f^{\text{pert}} - \boldsymbol{\xi}_f^{\text{nom}}\|_{\mathbf{R}_{[3],(t_f,t_0)}}},
 \end{equation}
and the absolute value of the difference between the final perturbed apoapsis radius and the nominal apoapsis radius normalized by this quantity for the perturbation along the maximal stretching direction are shown in \cref{fig:qcgt_mag}:
\begin{equation}
	\frac{|r_a^{\text{pert}} - r_a^{\text{nom}}|}{|r_a^{\text{pert}} - r_a^{\text{nom}}|_{\mathbf{R}_{[3],(t_f,t_0)}}}.
\end{equation}
For all HOCGTs, the perturbation along the eigenvector for the maximum eigenvalue from $(t_f, t_0)$ from the third-order HOCGT or augmented HOCGT induces the largest magnitude change in that coordinate. This indicates that the SS-HOPM solver does identify a direction of sensitivity for the nonlinear dynamics. Note that the y-axis is log scale, and there is still an order of magnitude difference in \cref{fig:hocgt_mag} and \cref{fig:scgt_mag} between the second-largest and largest relative objective change. For cases where the largest and second-largest relative objective changes are close, it may be prudent to keep more than one direction during DSTT construction (i.e. $l > 1$), depending on desired accuracy.

These results show that, for short propagation intervals, the direction of maximum stretching clearly varies both in time and between expansion orders throughout the trajectory. This indicates that a time-varying higher-order $\mathbf{R}_{[p]}$ direction could be beneficial for DSTT directionalization. However, as the propagation interval increases, the direction of maximum stretching generally stabilizes across time interval and order (although this is not the case for all scenarios and orders). Furthermore, we demonstrated that the HOCGTs and augmented HOCGTs correctly identify the direction that leads to the largest magnitude final perturbation in desired objective for the orthogonal directions considered. 

\section{Results}
\label{sec:results}

The previous sections presented multiple methods for finding maximum stretching directions using a higher-order Taylor series expansion of the dynamics with the goal of using these directions to construct more accurate DSTTs. In the following sections, we will compare DSTTs constructed using different $\mathbf{R}_{[p]}$ matrices in terms of the normalized Frobenius norm error between DSTTs and the STTs, perturbation propagation accuracy as a function of alignment with the $\mathbf{R}_{[p]}$ direction, and prediction accuracy for quantities of interest. All DSTT state perturbations are propagated using \cref{eqn:dstt_deterministic}, and all STT state perturbations are propagated using \cref{eqn:stt_pert}.

The following Taylor series approximations were compared:
\begin{enumerate}
	\item $\boldsymbol{\Phi}$, the STM; 
	\item $\phi^{[2]}$, the second-order STT;
	\item $\phi^{[3]}$, the third-order STT;
	\item 1-DSTT, the third-order DSTTs computed with the maximal eigendirection from the second-order CGT for both higher orders such that $\mathbf{R}_{[2]} = \mathbf{R}_{[3]} \in \mathbb{R}^{1 \times 7}$ and $y \in \mathbb{R}$;
	\item 3-DSTT, the third-order DSTTs computed using the eigenvectors corresponding to the largest three eigenvalues of the second-order CGT for both higher orders such that $\mathbf{R}_{[2]} = \mathbf{R}_{[3]} \in \mathbb{R}^{3 \times 7}$ and $\mathbf{y} \in \mathbb{R}^3$;
	\item 6-DSTT, the third-order DSTTs computed using the eigenvectors corresponding to the largest six eigenvalues of the second-order CGT for both higher orders such that $\mathbf{R}_{[2]} = \mathbf{R}_{[3]} \in \mathbb{R}^{6 \times 7}$ and $\mathbf{y} \in \mathbb{R}^6$ and $y \in \mathbb{R}$; 
	\item hoDSTT, the third-order DSTTs using the maximal eigenvector of the third- and fourth-order HOCGTs to directionalize the second- and third-order DSTTs respectively such that $\mathbf{R}_{[2]} \neq \mathbf{R}_{[3]}$ and $y \in \mathbb{R}$;
	\item sDSTT, the third-order DSTTs using the maximal eigenvector of the third- and fourth-order sCGTs (selecting position, velocity, and flight path angle) to directionalize the second- and third-order DSTTs respectively such that $\mathbf{R}_{[2]} \neq \mathbf{R}_{[3]}$ and $y \in \mathbb{R}$;
	\item $\varepsilon$-qDSTTs, the third-order DSTT using the maximal eigenvector of the third- and fourth-order qCGTs (with energy as the quantity of interest function) to directionalize the second- and third-order DSTTs respectively such that $\mathbf{R}_{[2]} \neq \mathbf{R}_{[3]}$ and $y \in \mathbb{R}$;
	\item and $r_a$-qDSTTs, the third-order DSTT using the maximal eigenvector of the third- and fourth-order qCGTs (with apoapsis radius as the quantity of interest function) to directionalize the second- and third-order DSTTs respectively such that $\mathbf{R}_{[2]} \neq \mathbf{R}_{[3]}$ and $y \in \mathbb{R}$. 
\end{enumerate}
All DSTTs are constructed using the full STM and DSTT approximations for the second- and third-order terms in the Taylor series expansion. 

\subsection{DSTT Approximation Accuracy}

To evaluate the accuracy of each DSTT construction, the normalized Frobenius norm error between DSTTs and STTs and perturbation propagation accuracy for different perturbation directions are computed. The baseline scenario described in \cref{sec:aerocapture} is used for the presented results, as the corresponding results for the high-velocity scenario are qualitatively similar.

\subsubsection{Normalized Frobenius Norm Error}

\begin{figure}[H]
    \centering
    \includegraphics{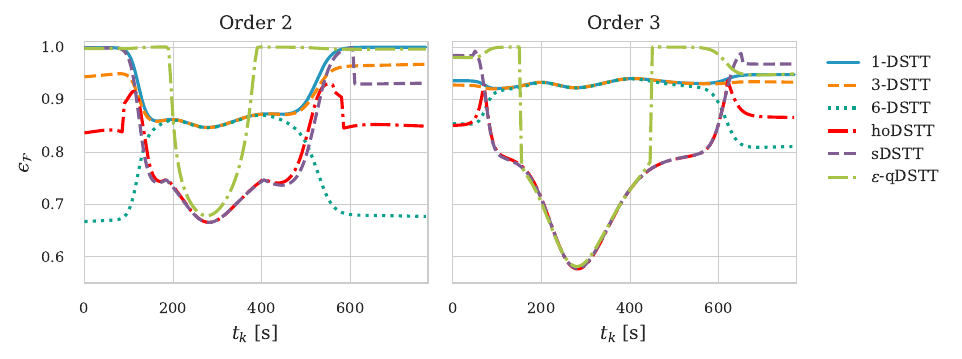}
    \caption{Normalized Frobenius norm error for different DSTT approximations mapping from $(t_{k+1}, t_k)$ for the baseline scenario.}
    \label{fig:norm_fro_norm}
\end{figure}

To determine how well a given DSTT approximates the original STT, the normalized Frobenius norm error is computed as:
\begin{equation}
	\epsilon_{\mathcal{F}} = \frac{\left|\left| \phi^{i, \gamma_1 \ldots \gamma_p} - \psi^i R_{[p]}^{\gamma_1} \cdots R_{[p]}^{\gamma_p}  \right|\right|_{\mathcal{F}}}{\left|\left|\phi^{i, \gamma_1 \ldots \gamma_p}\right|\right|_{\mathcal{F}}},
\end{equation}
where time indices are omitted for clarity. The Frobenius norm is the square root of the sum of the squared entries of the tensor. Lower error indicates more similarity  between the DSTT and STT. \cref{fig:norm_fro_norm} shows the normalized Frobenius norm error for various DSTT methods to propagate over the time domain $(t_{k+1}, t_k)$, where $\Delta t = t_{k+1} - t_k = 5$ seconds for the baseline scenario. The results for the high velocity scenario are similar to the results presented for the baseline scenario, so they are omitted for brevity. As apoapsis radius is only a function of the final state, the apoapsis radius qDSTT cannot be compared for this time domain.

The large initial errors for most methods indicate that the DSTTs are inaccurate reconstructions of the STTs for approximately the first 100 and final 200 seconds of the trajectory. In this region, the STTs near zero, indicating that there is low nonlinearity. The following discussions will focus on accuracy near maximum dynamic pressure (see \cref{fig:dynamic_pressure}) as the dynamics exhibit the most nonlinearity in this region. 

\begin{figure}[hbt]
	\centering
	\includegraphics{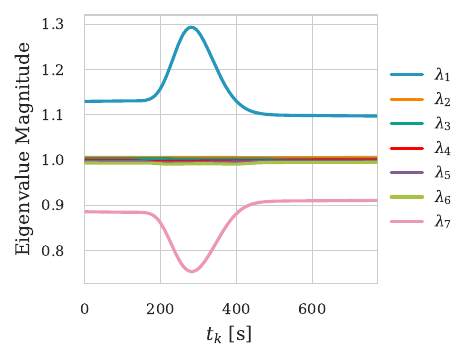}
  \caption{Eigenvalue magnitudes for $(t_{k+1}, t_k)$ for second-order CGT.}
  \label{fig:eval_mag}
\end{figure}

When evaluating DSTTs constructed with varying $\mathbf{y}$ dimensions (the 1-DSTT, 3-DSTT, and 6-DSTT), the initial error (between 0 and approximately 100 seconds) is improved when increasing the dimension of $\mathbf{y}$. However, the 1-DSTT, 3-DSTT, and 6-DSTT are equivalent from approximately 200 to 400 seconds, which corresponds to peak dynamic pressure (see \cref{fig:dynamic_pressure}). Accuracy is critical during this region, where small error in DSTT perturbation propagation during this region could compound and greatly impact terminal perturbation propagation accuracy. While the 6-DSTT is the most accurate of these DSTTs, there is little benefit to using a DSTT with this large of a latent dimension over the regular STT. Tensor contraction operations are still required to propagate a perturbation, rather than only matrix-vector multiplication for DSTTs with a single latent dimension (see \cref{eqn:dstt_deterministic}). In \cref{fig:eval_mag}, the eigenvalues over $(t_{k+1}, t_k)$ for the second-order CGT clearly show that $\lambda_1$ is between 10\% and 30\% greater than the other eigenvalues for all times. This indicates that a single dimension for $y$ is sufficient to capture the dominant linear effects of the dynamics, and adding additional latent dimensions in the original DSTT formulation provides only marginal improvement in approximation accuracy.

However, \cref{fig:norm_fro_norm} shows marked improvement when using the DSTTs constructed with the HOCGT, sCGT, and qCGT directions. The hoDSTT's error is lower than the original DSTTs (1-DSTT, 3-DSTT, and 6-DSTT) for the region of peak dynamic pressure. This demonstrates that for problems where the second-order and higher-order maximum stretching directions are not well aligned, there is a benefit to using the higher-order directions to construct the DSTTs. When constructing a second-order DSTT, using the maximum stretching direction of the third-order HOCGT, rather than the maximum stretching direction for the linear dynamics (second-order CGT maximal eigenvector), results in a DSTT which preserves more of the original STT. 

The sDSTT and $\varepsilon$-qDSTT have inferior error to the hoDSTT, especially for the initial and final times. While this indicates some accuracy loss during the initial and final time periods, these DSTTs preserve the high accuracy during the critical maximum dynamic pressure region and outperform the original DSTTs in that region. Because maximum dynamic pressure has a great impact on $\boldsymbol{\xi}$ and $\varepsilon$, it is more important for sDSTTs and $\varepsilon$-qDSTTs to retain accuracy during this period. 

In conclusion, using a higher-order direction with a single latent dimension improves DSTT approximation accuracy measured by the Frobenius norm, compared to employing more lower-order directions in a higher-dimensional latent space. These results suggest that the choice of projection direction is more important than the number of retained modes when constructing accurate DSTTs. Although DSTTs constructed from the proposed directions reduce Frobenius norm error relative to the original DSTT formulation, the remaining error is still greater than 50\%. Although Frobenius norm error characterizes how well the DSTTs approximate the original tensors, it is not directly correlated with perturbation propagation accuracy. For performance analysis applications, propagation accuracy is ultimately more relevant than Frobenius norm accuracy; this is examined in the following section.

\subsubsection{DSTT Perturbation Propagation Accuracy}

To investigate the impact of initial perturbation direction on the accuracy of DSTT propagation, we will explore prediction accuracy when propagating initial perturbation of varying directions. We will show that hoDSTTs are more accurate for perturbations closer to the maximal nonlinear stretching direction. This occurs because the scalar latent variable magnitude $|y| = \left| R_{[2]}^{i} \delta x_0^i\right|$ is maximized when $\mathbf{R}_{[2]}$ is parallel to $\delta \mathbf{x}_0$. Therefore, the hoDSTTs are more accurate for perturbations in directions near $\mathbf{R}$ such that $y$ is greater in magnitude, as can be seen from \cref{eqn:dstt_deterministic}. This fact is why the maximal linear stretching direction was originally chosen as the DSTT directionalization direction: it preserves the STT terms in a direction known to be nonlinear, where the STM will perform poorly.  

\begin{figure}[hbt]
	\centering
	\includegraphics{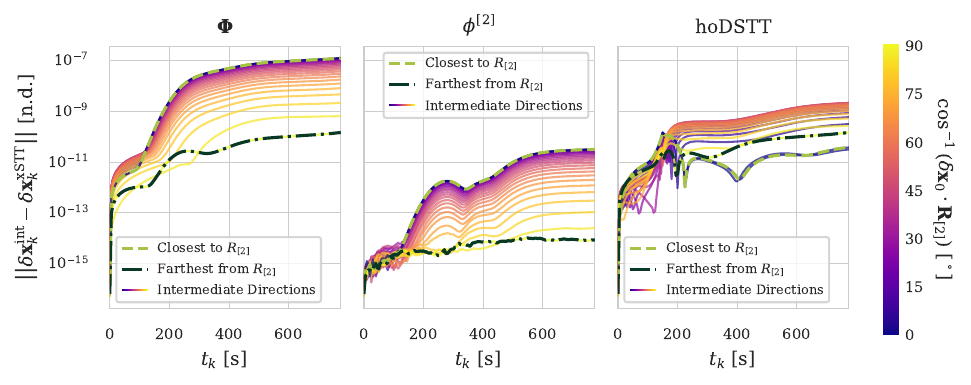}
  \caption{Perturbation propagation accuracy per perturbation magnitude along the $\mathbf{R}_{[2]}$ direction for the baseline scenario.}
  \label{fig:dstt_direction}
\end{figure}

To demonstrate the efficacy of using HOCGTs (and by proxy sCGTs and qCGTs) to directionalize DSTTs, the following study was conducted for the second-order scalar latent dimension hoDSTT. Given the maximal eigenvector of the third-order HOCGT over the entire timespan ($t_f, t_0$), $\mathbf{R}_{[2]}$, and a vector orthogonal to it, $\mathbf{u}$, such that $\mathbf{R}_{[2]}\cdot \mathbf{u} = 0$, we find vectors in directions $\delta \mathbf{x}_0 \propto \cos(\kappa) \mathbf{R}_{[2]} + \sin(\kappa) \mathbf{u}$ for $\kappa \in [0, 90]$ degrees. The orthogonal vector $\mathbf{u}$ is chosen using the same QR decomposition methodology described in \cref{sec:hocgt_comp}.\footnote{The QR decomposition results in six vectors orthogonal to $\mathbf{R}_{[2]}$. The results presented in this section hold regardless of which of the six vectors is selected as $\mathbf{u}$.} Low $\kappa$ angles correspond to directions near $\mathbf{R}_{[2]}$. \cref{fig:dstt_direction} shows the norm of Taylor series perturbation propagation error from the integrated solution for the baseline scenario for 25 angles $\kappa \in [0, 90]$ degrees, where each initial perturbation direction is scaled to have a nondimensional initial magnitude of $5 \times 10^{-6}$. Note that these results use the second-order STT and hoDSTT, while other studies in this paper use the third-order hoDSTT. 

The STM performs worse for directions near $\mathbf{R}_{[2]}$, but is more accurate for perturbations near-orthogonal to $\mathbf{R}_{[2]}$. This is consistent with the fact that a perturbation along the $\mathbf{R}_{[2]}$ direction should lead to the maximum final magnitude state perturbation (for linear and second-order dynamics). The second-order STT errors are near machine precision for the first 175 second of the trajectory for all directions, but the error increases after this time. The largest magnitude errors again correspond to directions closer to $\mathbf{R}_{[2]}$. When using a second-order hoDSTT directionalized along the third-order HOCGT maximal eigenvector ($\mathbf{R}_{[2]}$), the error is lower than the STM for directions far from $\mathbf{R}_{[2]}$. Interestingly, the error is also lower than the STM error for directions close to $\mathbf{R}_{[2]}$. The directions between 30 and 60 degrees from $\mathbf{R}_{[2]}$ have the largest error for the hoDSTT method, but the error for these perturbations is still lower than the error for the same perturbations propagated with only the STM.

Because the magnitude of $y$, $|y| = \left| R_{[2]}^{i} \delta x_0^i\right|$, is maximized for $\delta \mathbf{x}_0$ directions near parallel $\mathbf{R}_{[2]}$, even a single latent dimension hoDSTT improves perturbation propagation accuracy for these nonlinear directions. The quantity $|y|$ is minimized (near zero) for $\delta \mathbf{x}_0$ directions near orthogonal to $\mathbf{R}_{[2]}$, but because these directions do no excite nonlinearity in the dynamics, they do not degrade the accuracy of the DSTT.

\subsection{Quantity of Interest Propagation}

To compare DSTT perturbation propagation performance for key quantities of interest, 10,000 $\delta \xvec_0$ samples were taken from the Gaussian distribution in \cref{tab:init_cov} and propagated through each Taylor series approximation. The following subsections evaluate propagation accuracy for terminal apoapsis radius and energy and a time history of energy. Results for terminal apoapsis radius and energy prediction are only included for the baseline scenario outlined in \cref{sec:aerocapture}, as the results from the high velocity scenario are similar. Results from both the baseline and high velocity scenarios are compared for energy time history propagation. Although DSTTs can be used to propagate the moments of a Gaussian distribution~\cite{boone_directional_2023}, these results are omitted from this paper as they are consistent with the deterministic perturbation propagation results.

\subsubsection{Terminal Apoapsis Radius and Energy Prediction}

\begin{figure}[htb]
	\centering
	\includegraphics{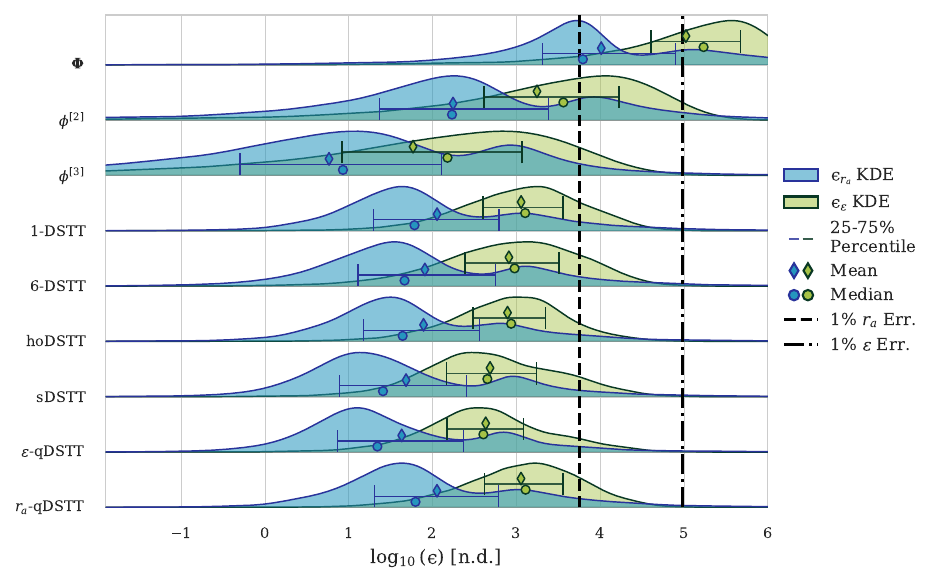}
  \caption{Ridge plot of the standard logarithm of terminal apoapsis radius and energy error for the baseline scenario.}
  \label{fig:energy_apoapsis_error}
\end{figure}

\cref{fig:energy_apoapsis_error} presents KDE fits to the standard logarithm of apoapsis and terminal energy error for each propagation methods. The procedure to obtain these metrics is described in \cref{sec:stt_example}. Note that all DSTT expansions are up to third-order, while the first- through third-order STT results are shown.

For terminal energy, the STM has the greatest error values and relatively small spread in the data, while the third-order STT errors are the lowest magnitude and highest spread. There is little performance difference for DSTTs with different latent dimensions (the 1-DSTT and 6-DSTT), with only slightly lower median error when increasing the latent dimension from one to six. This again verifies the fact that for systems where the largest eigenvalue magnitude is much greater than the others, there is little improvement in prediction accuracy when adding more of the second-order CGT bases to the DSTT. For energy, there is also little difference between the DSTTs constructed with eigenvectors from the second-order CGT (1-DSTT and 6-DSTT) and the DSTT constructed with the dominant eigenpair from the third- and fourth-order CGTs (hoDSTTs). However, when using the maximal eigenpairs of the augmented higher-order sCGT and $\varepsilon$-qCGTs to directionalize the DSTTs, median energy error decreases by half an order of magnitude, along with reduction in the 75$^\text{th}$ percentile value. This demonstrates that using an augmented HOCGT parameterized by either the constituent states (in this case, $r$, $V$, and $\gamma$) in a quantity of interest function or that function itself can lead to improved DSTT propagation performance in that quantity of interest. 

However, this finding is not true when looking at the apoapsis radius error. While the same trends as for the energy error hold for the 1-DSTT, 6-DSTT, hoDSTT, sDSTT, and $\varepsilon$-qDSTT, the $r_a$-qDSTT performs similarly to the 1-DSTT in predicting terminal apoapsis. While the $\varepsilon$-qDSTT was more accurate at predicting energy than DSTTs constructed using other methods, the $r_a$-qDSTT is not more accurate at predicting apoapsis radius. This is likely because apoapsis radius is a poorly-conditioned function for quantifying aerocapture performance. Because of the discontinuity in semi-major axis when the trajectory transitions from hyperbolic to elliptical, apoapsis radius exhibits asymptotic behavior mid-flight (see \cref{eqn:apoapsis_radius}). This discontinuity is not present for energy, which monotonically decreases throughout an aerocapture trajectory (see \cref{eqn:specific_energy}). When computing the apoapsis radius partials to obtain the qCGTs, the discontinuity induces numerical issues, which can in turn result in inaccurate DSTTs. However, the sDSTT that selects the states used in the apoapsis radius function \cref{eqn:apoapsis_radius}, $\boldsymbol{\xi} = \left[ r, V, \gamma \right]^{\top}$, is more accurate in apoapsis radius error than the other DSTTs. For applications requiring apoapsis radius prediction, the sDSTT approach (constructing a sCGT for the states in the quantity of interest function) is more accurate than the qDSTT approach (constructing a qCGT using the quantity of interest function). 

DSTTs constructed using nonlinear stretching directions for subsets of the state and numerically-stable functions of the state can be more accurate at propagating quantities of interest than DSTTs constructed from the second-order CGT maximal eigenpair. In specific, DSTTs constructed using a single direction leveraging the higher-order dynamics are at least as accurate as DSTTs constructed using six directions based only on linear stretching. However, if the quantity of interest function is discontinuous or has poorly-defined partials, as with apoapsis radius, the DSTTs are not necessarily more accurate. In these cases, constructing an sDSTT with the constituent state variables in the nonlinear quantity of interest function results in improved performance.

\subsubsection{Energy Time History Propagation}

\begin{figure}[hbt]
	\centering
	\begin{subfigure}[b]{0.48\textwidth}
         \centering
         \includegraphics[width=\textwidth]{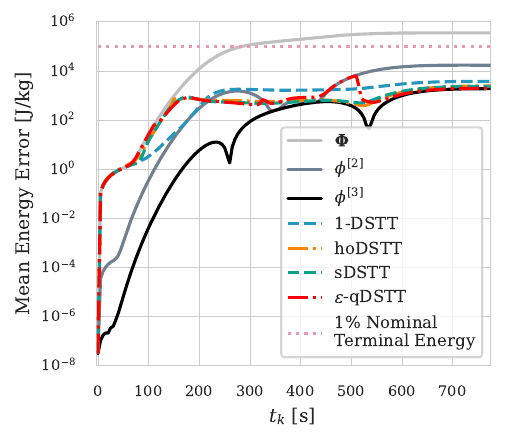}
         \caption{Baseline Scenario.}
         \label{fig:eng_err_baseline}
     \end{subfigure}
     \hfill
     \begin{subfigure}[b]{0.48\textwidth}
         \centering
         \includegraphics[width=\textwidth]{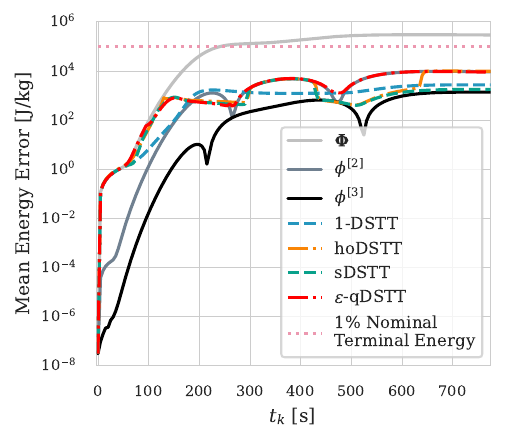}
         \caption{High Velocity Scenario.}
         \label{fig:eng_err_fast}
     \end{subfigure}
     \caption{Mean energy error from 10,000-case Monte Carlo simulation.}
     \label{fig:energy_error}
\end{figure}

The same 10,000 $\delta \xvec_0$ samples are propagated through Taylor series methods from $(t_k, t_0)$ and the energy computed for the true integrated perturbed trajectory is compared with the energy computed with the Taylor series approximation. The mean of the error between the analytical solution and the integrated solution for both the baseline and high-velocity scenarios are shown in \cref{fig:energy_error}. For all times for both scenarios, all higher-order solutions result in errors lower  than 1\% of the nominal terminal energy. All DSTTs perform similarly to the STM for approximately the first 100 seconds of flight. Despite containing third-order terms, the DSTTs do not achieve near third-order accuracy for energy propagation for propagation intervals of less than 100 seconds. This could indicate that the transient change in dominant direction during this initial period, as seen in \cref{fig:hocgt_t0tk}, results in poor DSTT propagation accuracy. 

For both scenarios, the 1-DSTT performs the best of all DSTTs from 100 to 200 seconds, where all augmented DSTTs still perform similarly to the STM. As can been seen in \cref{fig:hocgt_t0tk}, the second-order CGT dominant eigenvector smoothly stabilizes to its steady state value over this time interval, while the higher-order directions vary more drastically. This could again indicate that a smooth transition in DSTT direction, or a direction more similar to the final steady state value, result in more accurate DSTTs for  $\Delta t = t_k - t_0 < 200$ s for propagation from $(t_k, t_0)$.

The results for the remainder of the trajectory differ slightly between the two scenarios. For the baseline trajectory, the third-order hoDSTTs, sDSTTs, and $\varepsilon$-qDSTTs generally result in lower error than the 1-DSTT and outperform the second-order STT. The only time interval where this is not true is from 450 to 500 seconds, where the third-order $\varepsilon$-qDSTT has similar error to the second-order DSTT. For this time regime, the $\varepsilon$-qDSTT is not retaining any third-order information. The $\varepsilon$-qDSTT is specifically directionalized along the maximum stretching direction for energy change on each $(t_k, t_0)$ interval. In contrast, the sDSTT is constructed along the direction that maximizes the nondimensional magnitude of vector $\boldsymbol{\xi} = \left[ r, V, \gamma \right]^{\top}$, some of the the constituent states in energy expression (see \cref{eqn:specific_energy}). These results show that using DSTTs constructed from (1) a sCGT of the constituent states in a quantity of interest function or (2) a qCGT for that quantity of interest function can provide comparable performance, although not necessarily for all time intervals. 

For the high velocity scenario, many of the same trends are present as in the baseline case for the final 600 seconds of flight. However, there are periods of time where both the third-order hoDSTT, sDSTT, and $\varepsilon$-qDSTT are equivalent to the second-order STT. In fact, the $\varepsilon$-qDSTT performs similarly to the second-order STT throughout most of flight. The regions of time where the augmented DSTTs perform poorly correspond to regions where the fourth-order HOCGT and augmented HOCGT maximal eigenpairs vary, as shown in \cref{fig:hocgt_t0tk}. These directions are used to construct the third-order DSTTs. For these time regimes, using the linear stretching direction for the third-order DSTTs (as is done to construct the 1-DSTT) results in lower propagation error than using higher-order stretching directions. 

This illustrates a potential drawback of using tensor eigenpairs to construct DSTTs: while the SS-HOPM solver is guaranteed to converge to an eigenpair of the system, it may not always converge to the maximal eigenpair (see discussion in \cref{sec:hocgts}). Despite comparing multiple solutions from SS-HOPM to find the maximal eigenpair, this method could still fail to return the true maximum. For the regions where the third-order augmented DSTTs perform similarly to the second-order STTs, there is significant variation in the direction associated with the largest eigenvalue (see \cref{fig:hocgt_t0tk}), which could indicate that multiple eigenvalues with similar magnitudes exist.

 For much of the time regimes presented here, the novel DSTTs presented in this paper perform worse than the original DSTTs for the high velocity scenario. While all third-order DSTTs result in error less than or equal to the second-order STT for most of the trajectory, they are not better than the second-order STT. However, the third-order DSTTs still contain significantly fewer terms than the second-order STT (the third-order single latent dimension DSTTs contain 14 terms to map third-order information, while the second-order STT contains 343 terms). Overall, the augmented DSTTs presented in this work result in a compact and computationally efficient alternative to STTs for higher-order perturbation propagation.

\section{Conclusions} 
\label{sec:conclusions}

This paper derived new methods for constructing reduced-dimension DSTTs for aerocapture and demonstrated that STTs and DSTTs are effective tools for this highly nonlinear, nonconservative aerocapture dynamical system. Previous work relied on finding the direction of maximum linear stretching to reduce the dimension of the higher-order STTs; in this work, the tensor eigenpairs of HOCGTs and novel augmented HOCGTs were used to construct reduced-dimension DSTTs. The tensor eigenvector of the maximum eigenvalue of a HOCGT is the direction which leads to the maximum final state magnitude perturbation due to the nonlinear dynamics. This idea was extended to develop sCGTs, whose maximal eigenpair determines the direction which leads to the maximum final magnitude perturbation in the nonlinear dynamics for a selected subset of the full state, and qCGTs, whose maximal eigenpair determines the direction which leads to the maximum final magnitude perturbation through the nonlinear dynamics for a nonlinear function of the state. The latter two augmented HOCGT formulations recognize that certain states matter more than others for aerocapture applications, which can be leveraged to construct a DSTT that is most accurate for a specific application. 

DSTTs constructed using these dynamics-informed higher-order directions were shown to have lower Frobenius norm error relative to the STTs. In addition, we showed that DSTTs constructed with these novel directions for a single latent dimension can be more accurate for perturbation propagation than DSTTs constructed with the original DSTT linear stretching directions and larger latent dimensions. The novel DSTTs were also more accurate for propagating key quantities of interest than the original DSTTs. 

The tensor eigenpairs of augmented HOCGTs can be applied to any nonlinear dynamical system to construct DSTTs while considering nonlinear directions of sensitivity. Other applications for augmented HOCGT tensor eigenpairs include Gaussian mixture model splitting to enable nonlinear, non-Gaussian state distribution propagation, which can be used for more accurate filtering. Computationally efficient DSTT-based predictions could also be incorporated into guidance algorithms. 

\section*{Funding Sources}
This work was supported by NASA Space Grant Technology Research Fellowship grant number 80NSSC23K1227. 

\section*{Acknowledgments}
G. Calkins would like to thank Oliver Boodram and Dillon Waxman for their discussions and review of the tensor algebra in this paper and Jackson Kulik for his insight on tensor eigenpairs. Artificial intelligence tools were used for LaTeX equation editing, Python coding assistance in figure generation, and language editing to improve clarity and readability.

\bibliography{references}

\end{document}